\documentclass[submitting]{nst}

\usepackage{subfigure,dcolumn}
\usepackage[T2A,T1]{fontenc}
\usepackage[english]{babel}
\usepackage{subfigure}
\usepackage{booktabs}

\begin{document}
\title{Screener3D: a Gaseous Time Projection Chamber for Ultra-low Radioactive Material Screening}
\author{Haiyan Du}
\affiliation{INPAC; Shanghai Laboratory for Particle Physics and Cosmology;
	Key Laboratory for Particle Astrophysics and Cosmology (MOE), \\
	School of Physics and Astronomy, Shanghai Jiao Tong University, Shanghai {\rm 200240}, China}

\author{Chengbo Du}
\affiliation{Yalong River Hydropower Development Company, Ltd., 288 Shuanglin Road, Chengdu 610051, China}

\author{Karl Giboni}
\author{Ke Han}
\email[Corresponding author, ]{ke.han@sjtu.edu.cn}
\affiliation{INPAC; Shanghai Laboratory for Particle Physics and Cosmology;
 Key Laboratory for Particle Astrophysics and Cosmology (MOE), \\
 School of Physics and Astronomy, Shanghai Jiao Tong University, Shanghai {\rm 200240}, China}

\author{Shengming He}
\author{Liqiang Liu}
\affiliation{Yalong River Hydropower Development Company, Ltd., 288 Shuanglin Road, Chengdu 610051, China}
\author{Yue Meng}
\affiliation{INPAC; Shanghai Laboratory for Particle Physics and Cosmology;
 Key Laboratory for Particle Astrophysics and Cosmology (MOE), \\
 School of Physics and Astronomy, Shanghai Jiao Tong University, Shanghai {\rm 200240}, China}

\author{Shaobo Wang}
\affiliation{INPAC; Shanghai Laboratory for Particle Physics and Cosmology;
 Key Laboratory for Particle Astrophysics and Cosmology (MOE), \\
 School of Physics and Astronomy, Shanghai Jiao Tong University, Shanghai {\rm 200240}, China}
\affiliation{SPEIT~(SJTU-ParisTech Elite Institute of Technology), Shanghai Jiao Tong University, Shanghai, {\rm 200240}, China}
\author{Tao Zhang}
\author{Li Zhao}
\affiliation{INPAC; Shanghai Laboratory for Particle Physics and Cosmology;
 Key Laboratory for Particle Astrophysics and Cosmology (MOE), \\
 School of Physics and Astronomy, Shanghai Jiao Tong University, Shanghai {\rm 200240}, China}
\author{Jifang Zhou}
\affiliation{Yalong River Hydropower Development Company, Ltd., 288 Shuanglin Road, Chengdu 610051, China}

\begin{abstract}{
In experiments searching for rare signals, background events from the detector itself are some of the major factors limiting search sensitivity.
Screening for ultra-low radioactive detector materials is becoming ever more essential.
We propose to develop a gaseous time projection chamber (TPC)  with a Micromegas readout for radio screening.
The TPC records three-dimensional trajectories of charged particles emitted from a flat sample placed in the active volume of the detector.
The detector can distinguish the origin of an event and identify the particle types with information from trajectories, which significantly increases the screening sensitivity.
For $\alpha$ particles from the sample surface, we observe that our proposed detector can reach a sensitivity higher than 100~$\mu$Bq$\cdot$m$^{-2}$ within two days.
}\end{abstract}

\maketitle
\section{Introduction}
Experiments searching for neutrinoless double beta decay~\cite{DellOro:2016tmg} and dark matter direct-detection signals~\cite{Roszkowski:2017nbc,Liu:2017drf} have extremely low signal rates, if any at all.
The background rate recorded by the detector is essential to the search sensitivity. 
In underground laboratories such as the China Jinping Underground Laboratory (CJPL)~\cite{Cheng:2018lcf}, the influence of cosmic rays is reduced by orders of magnitude and is no longer dominant.
Background events result almost entirely from the detector itself and the surrounding lab environment.
Therefore, extensive material screening campaigns to select materials with low radioactivities have been conducted  by many experiments (for example, ~\cite{Wang:2016eud,Jiang:2018pic,Abgrall:2016cct,Akerib:2014rda,Leonard:2007uv,Alessandria:2011vj}). 
The majority of efforts emphasize the bulk radioactivities of materials using screening techniques such as $\gamma$-ray measurement with high purity germanium (HPGe) detectors, inductively coupled plasma mass spectrometry (ICP-MS), and neutron activation analysis (NAA).
Note that these techniques have significantly wider applications than only screening low-radioactive materials  (for example, ~\cite{HPGe_application,ICP-MS_application}). 

Surface radioactivity often differs from those in the bulk and may significantly contribute to the background budget. 
The surface of detector components may become contaminated with additional radioactivity through machining, handling, or exposure to air.
When directly facing a detector active volume, $\alpha$ or $\beta$ particles emitted from the surface may directly introduce  background events. 
The surface background is particularly  critical for modular detectors where large surface areas face detector modules.
For example, $\alpha$ events from the supporting structure and surface of a detector are the most prominent source in the Cryogenic Underground Observatory for Rare Events (CUORE)  bolometer array~\cite{Alduino:2016vtd}.
Even if surface contaminations are far from detectors, radioactive impurities may emanate from the surface and travel to the detector via the circulation of the liquid or gas detector medium.
For example, in the PandaX-4T experiment, $^{222}$Rn from the detector and inner surface of the circulation pipe is a major background source in the search for dark matter~\cite{Zhang:2018xdp}.

Surface radioactivity is often measured using semiconductor detectors, scintillator arrays, gas ionization counters, etc.
The detection areas of commercial silicon semiconductor detectors are frequently smaller than 30~cm$^2$ (e.g.~\cite{OrtecWebsite}).
An example of a scintillator array is BiPo-3, which utilizes 40 scintillator modules for a total screening area of 3.6~m$^2$~\cite{BiPo-3}. 
The background rate achieved by the BiPo-3 detector is 0.9~$\mu$Bq$\cdot$m$^{-2}$ for $^{208}$Ti and 1~$\mu$Bq$\cdot$m$^{-2}$ for $^{214}$Bi~\cite{BiPo-3,BiPo-3_background}.
The extremely sensitive detector requires dedicated support; therefore, the cost for measurement is high. 
Gas ionization counters can have large active areas and particle identification capability using pulse shape analysis.
Ultralo 1800, offered by XIA~\cite{XIAWebsite}, is one such detector with an $\alpha$ background of approximately 280 $\mu$Bq$\cdot$m$^{-2}$.
However, it fully utilize the three-dimensional track recording capabilities of gas detectors. 
Proposals such as BetaCage~\cite{BetaCage} and $\mu$ time projection chamber ($\mu$-TPC)~\cite{upic} are no longer active or are in the early stage of development. 
Our proposal differs from $\mu$-TPC in terms of sample placement, readout modules, and operating gas medium.

We propose to construct a large-area, high-efficiency, and low-cost gaseous TPC to measure surface contaminations, which we call Screener3D. 
Screener3D will be able to measure the energy and trajectories of $\alpha$ and $\beta$ particles with position-sensitive readout modules of the Micro-MEsh Gaseous Structure (Micromegas~\cite{MM}).
In a typical argon TPC operating at atmospheric pressure, $\alpha$ particles of 10~MeV travel approximately 10~cm in an almost straight line.
Trajectories of $\beta$ particles of the same energy are significantly longer and meander.
Along the particle trajectories, the gas medium is ionized. 
The number and initial position of ionization electrons carry the three-dimensional trajectory information of the original particle.
When combining energy and trajectory information, we can better distinguish signals from the background and identify the origin of the signals.
Samples are placed in the TPC and the emitted charged particles are recorded with high efficiency.
For $\alpha$ particles, in particular, the energy can be measured more accurately without energy loss in entrance windows. 
In this paper, we provide an overview of the proposed design and sensitivity studies using simulations.

\section{Design overview}\label{sec:design}
The central part of Screener3D is a low-background, large-area, and high-granularity gaseous TPC.
This type of gaseous TPC is extensively used in nuclear and particle physics (e.g.~\cite{CAT-TPC,PandaX-III})
All the materials used to build the TPC will have low radioactivities to control the background contributions from the detector itself.
The designed readout area, which determines the largest sample area that can be measured, is approximately  2000~cm$^2$.
The Micromegas readout will be based on 3~mm wide strips, a granularity sufficient for $\alpha$ and $\beta$ tracks in the order of 10~cm.
In addition to the key detector performance specifications, Screener3D must be stable over an extended period. 
The long-term stability is maintained with a gas circulation and purification system and a real-time monitoring system.
With the auxiliary systems, we can measure samples for a prolonged period of up to weeks to increase  the measurement sensitivity.
We describe the main components of Screener3D and material screening plans in this section.

\subsection{Gaseous TPC with Micromegas}
\begin{figure}[tb]
 %\vspace{-0.01cm}
 \centering
 \includegraphics[width=\columnwidth]{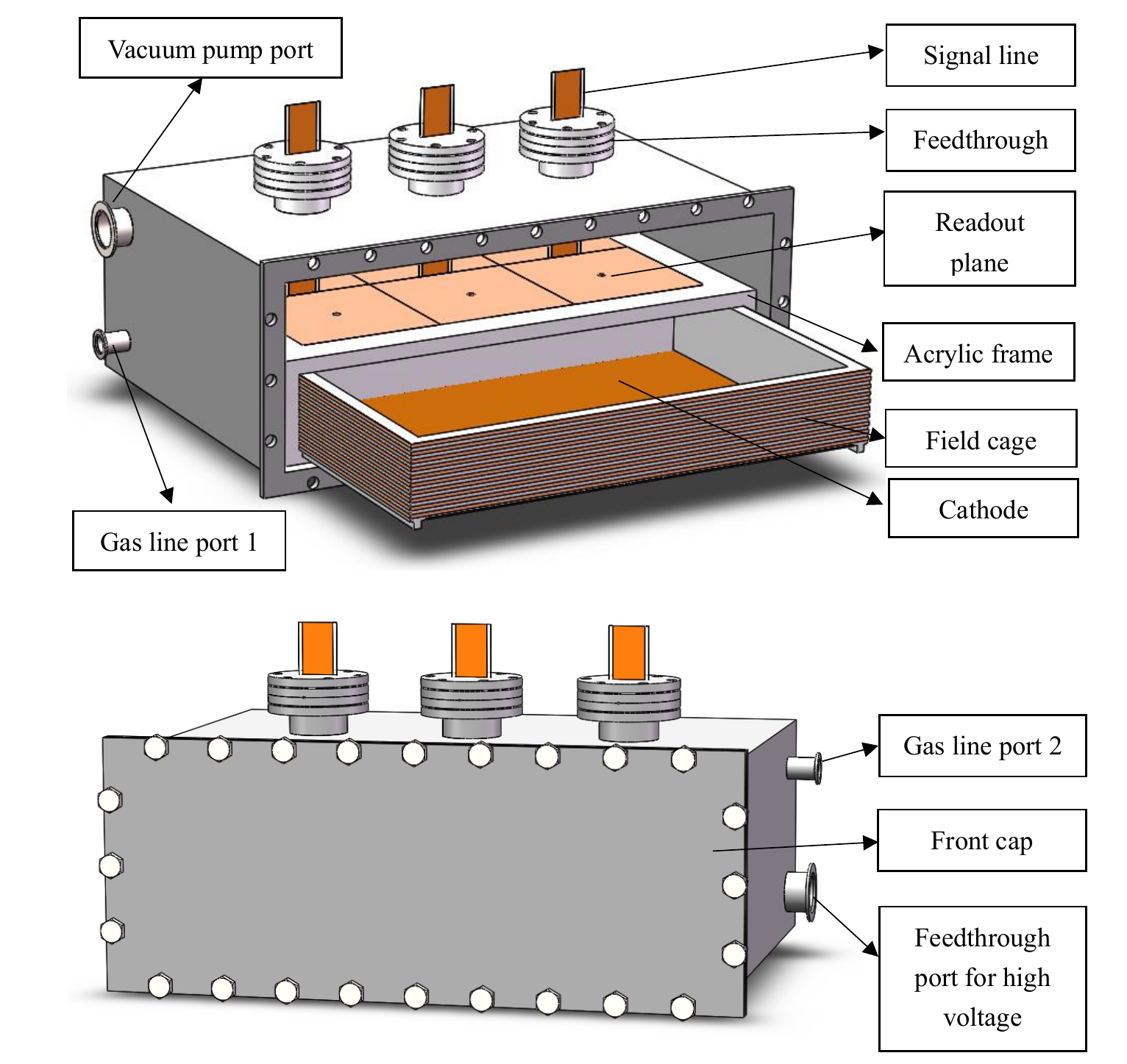}
 \caption{Schematic design of Screener3D with the main components labeled. (Top) The inside view with field cage drawer pulled out. Samples are placed at the bottom of the drawer (on the cathode). (Bottom) The detector with front cap closed and ready for measurement.}
 \label{fig:DrawingTPC}
\end{figure}
The Screener3D TPC primarily comprises a readout plane with Micromegas modules, a cathode, a field cage, a gas medium, and an outer vessel (Fig.~\ref{fig:DrawingTPC}). 
The rectangular readout plane on the top collects drift electrons after amplification via avalanche.
The cathode is at the bottom and provides a negative high voltage. 
The field cage connects the readout plane and cathode mechanically and aids in maintaining a uniform drift electric field in the active volume, which is the cuboid space enclosed by the aforementioned components.
The active volume is 60~cm long, 40~cm wide, and 10~cm high, which is filled with 1 bar of an argon and isobutane gas mixture during operation.
All the components are fixed inside the vessel with signal feedthroughs and various ports for gas circulation and pumping. 

During measurement, a thin slab sample, such as a sheet of metal, plastic film, or silicon wafer, would be placed on the cathode plane.
Radioactive contaminations on the top surface of the sample emit $\alpha$ and $\beta$ particles into the active volume.
We focus primarily on the identification of $\alpha$ particles from samples since the energy deposition in the unit distance is large and trajectory characteristics are easier to define. 
The typical $\alpha$ emissions emanate from uranium and thorium decay chains. 
Nuclides such as $^{232}$Th, $^{238}$U, $^{214}$Po, and $^{212}$Po emit $\alpha$ particles with energies of 4.0, 4.1, 7.7, and 8.8~ MeV, respectively.
A few other isotopes along the decay chains emit $\alpha$ particles in the energy range of 4 to 7~MeV.

The readout plane may consist of multiple Micromegas modules.
In our current design, six  modules of approximately $20\times20\,\rm{cm}^2$ are tiled together to form a total readout area of 2400~cm$^2$.
In Fig.~\ref{fig:DrawingTPC}, the modules, supported by an acrylic frame, are shown face-down.
For each module, the active area is split into $64\times64$ diamond-shaped pads, each of which has a diagonal length of 3~mm.
The pads are inter-connected in the vertical or horizontal direction and form the so-called X or Y readout strips. 
With 64 X and 64 Y strips, the total number of electronics channels is 128 per module.
Flat signal cables fabricated from a Kapton-based flexible printed circuit board (PCB) pass through the feedthroughs on the top of the vessel to send the signals to readout electronics (not shown in the figure) outside the TPC.
Micromegas modules are frequently biased at a few hundreds of volts and the voltage is also provided through the flat Kapton cable.

Fig.~\ref{fig:DrawingTPC} (top) also shows the field cage and cathode as a drawer pulled out from the acrylic frame. 
The sidewall of the field cage is created from 2~cm-thick acrylic plates with a Kapton-based flexible PCB attached to the outside.
There are uniformly spaced copper strips on the PCB, and the adjacent strips are connected with surface mount resistors to maintain a uniform electric field.
At the bottom of the drawer, a polished copper plate with a thickness of 0.2~cm is used as the cathode.
The cathode is electrically connected to the bottom strip of the field cage PCB. 
A uniform electric field is formed within the field cage when a negative high voltage is provided to the cathode. 
The entire setup is enclosed in an outer vessel created from 1~cm-thick stainless steel (SS).
For easy sample loading with the drawer-style field cage, the opening of the vessel is at the front (Fig.~\ref{fig:DrawingTPC} (bottom)).
Ports for gas lines, a vacuum pump, and feedthrough for the cathode high-voltage cable are added on the side of the vessel. 

Samples are placed on the cathode for measurement.
In our current design shown in Fig.~\ref{fig:DrawingTPC}, the front panel of the vessel is unmounted, the field cage is pulled out, and the sample can be placed on the cathode.
When closed again, we pump down the vessel to $\mathcal{O}(1)$ mPa before flushing argon gas in.
We expect the entire sample loading process to be accomplished in a period of an hour. 
Alternative designs, such as a stationary field cage with a slit opening on the front, are also under consideration to further streamline the sample loading process.

\begin{figure}[tb]
	%\vspace{-0.01cm}
	\centering
	\subfigbottomskip=5pt 
	\subfigcapskip=3pt 
	\subfigure[Top view of the detector showing potential on the X--Y plane]{
		\includegraphics[width=0.9\linewidth]{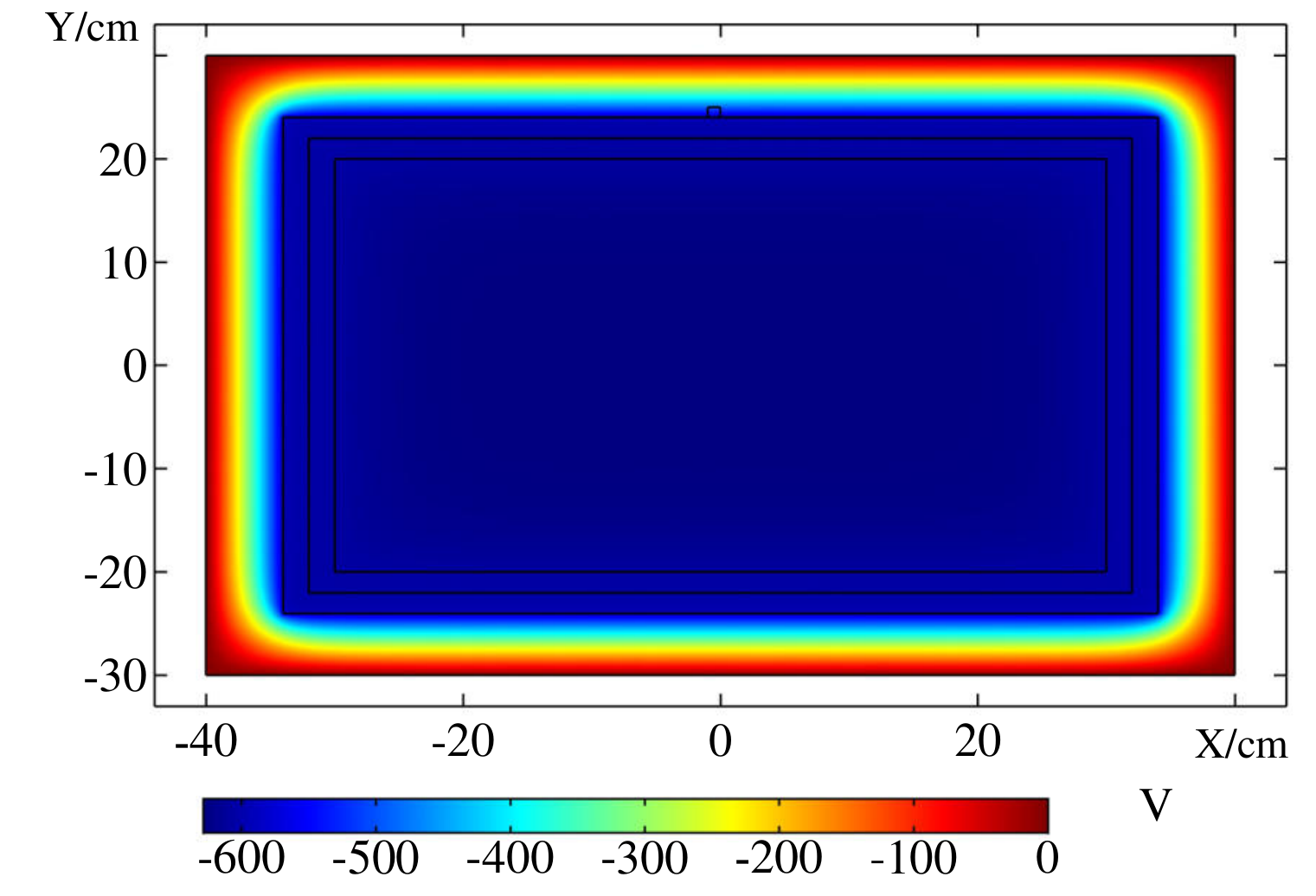}}
	\subfigure[Side view of the detector showing potential on the X--Z plane]{
		\includegraphics[width=0.9\linewidth]{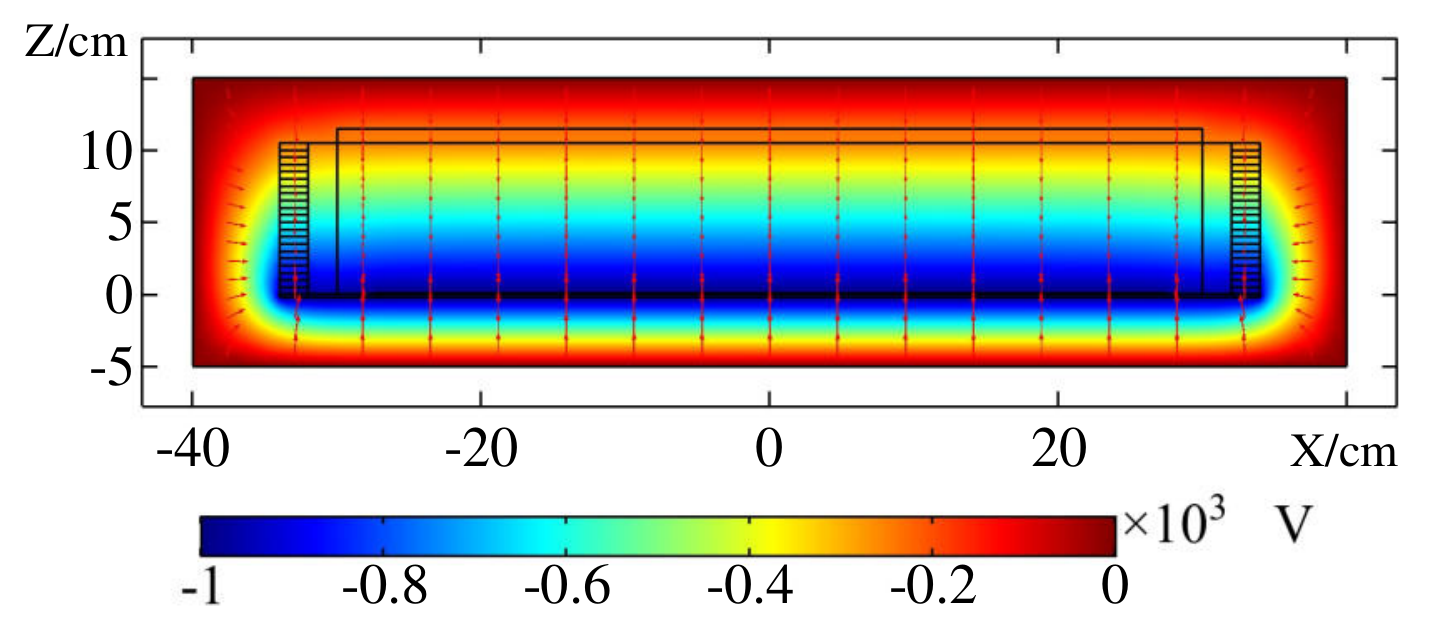}}
	\caption{Electric field simulation using COMSOL, which demonstrated minimal impact to the electric field of samples on the cathode. (a) Top view of the detector showing the potential on the X--Y plane. The innermost black rectangle shows the boundary of the sample, which had a slightly smaller footprint than the active volume of the TPC. (b) Side view of the detector showing the potential on the X--Z plane. Red lines with arrows indicate the electric field lines. We can observe that the field lines were in almost perfect vertical directions above the sample's footprint.}
	\label{Electric_field_simulation}
\end{figure}

When samples are placed in the active volume, the counting efficiency is high and the measurement of $\alpha$-particle energy is more accurate. 
An aspect of concern may be the disturbance of electric potential lines when samples are on the cathode.
We simulated  the effect of a thin Kapton sample using the finite element analysis software COMSOL Multiphysics ~\cite{COMSOLWebsite} (Fig.~\ref{Electric_field_simulation}).
The black wireframes in Fig.~\ref{Electric_field_simulation} represent the internal structure of the TPC, such as the field cage, cathode, and readout modules.
We supplied the cathode with a 1~kV negative voltage and the readout plane with a 250~V negative voltage.
The Kapton sample 54.4~cm long, 34.4~cm wide, and 0.1~cm thick is shown as the innermost rectangle in Fig.~\ref{Electric_field_simulation} (a).
Fig.~\ref{Electric_field_simulation} (b) shows the side view of field cage. 
The field lines inside the field cage exhibited minimal distortions. 
In the X--Z plane , the electric field in the X-direction was 3.6\% of that in the vertical Z-direction at 2~mm from the edge of the sample, while the value was less than 1\% at 2~cm from the edge of the sample.
The uniformity of the electric field was almost the same for a 0.1~cm-thick copper or silicon sample of the same size; thus, it was negligible for our measurement.

The distortion of the electric field at the edge of the sample became more severe with increased sample thickness. 
For a 1~cm-thick copper sample, the distortion became smaller than 1\% when measured at approximately 7 cm from the edge. 
Therefore, we can mitigate this problem using more aggressive fiducial cuts.
However, we should note that the main objective of Screener3D is to measure thin millimeter-scale samples.

\subsection{Gas system and slow monitoring system}\label{sec:auxiliary}
\begin{figure}[tb]
 %\vspace{-0.01cm}
 \centering
 \includegraphics[width=\columnwidth]{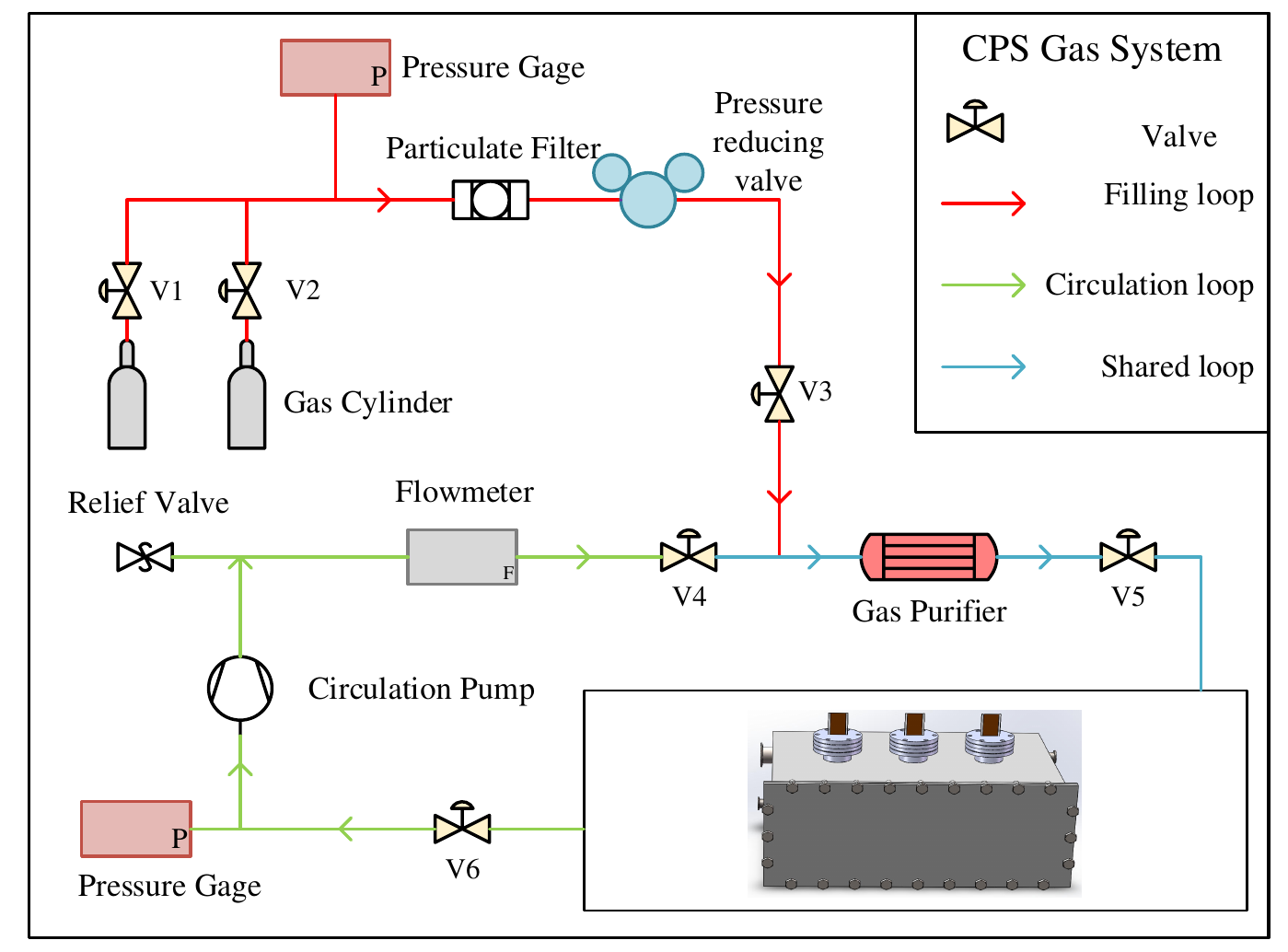}
\caption{Schematic diagram of the Screener3D gas system. Red arrows denote the filling loop. Green arrows denote the circulation loop. Blue arrows are the loops shared by the filling and circulation. After the pressure-reducing valve, the working pressure is limited to 1 bar.}
 \label{gas system}
\end{figure}

The energy response and resolution of gaseous TPC are highly sensitive to the purity of gas medium, particularly electronegative impurities such as oxygen and water.
The impurities may attract and absorb drift electrons in the active volume.
They may also reduce the gain in the avalanche gap of the Micromegas.
For the stability of the prolonged measurement period of Screener3D, a gas circulation and purification system (shown in Fig.~\ref{gas system}) are designed.
Before flowing into the TPC, gas mixtures travels through a set of gas purifiers. 
Given the different precision requirements of measurement, the purifier may be a charcoal-based absorber (primarily to absorb radio-impurities such as radon), a chemical-reaction-based getter (primarily to absorb electronegative impurities), or a combination of both.
A circulation loop is added for the continuous purification of the working gas for an extended measurement of low-radioactivity samples.

To quantify the long-term (in)stability of the detector, we monitor key operation parameters, such as ambient temperature, detector temperature, gas pressure, Micromegas bias voltage, leakage current, TPC drift voltage, and event rates .
We design and build a centralized slow monitoring system~\cite{slow_control} (SMS) with all the parameter values stored in a database and are accessible in real-time on a web browser. 
With pre-defined safety ranges of the key parameters, the SMS can also send alerts via email or phone text messages.
The SMS data would also enable us to reject or correct poor-quality data based on the variation of operating conditions.
The SMS system may also store sample information, such as area, material, and screening time to serve as a centralized catalog of measurements.

\subsection{Intrinsic background sources and mitigation}\label{sec:materials}
To improve the signal-to-background ratio in screening, we must minimize the radioactive background from Screener3D itself. 
The inner part of the Screener3D TPC will be built with materials of low radioactivity, screened using HPGe and/or mass spectrometer techniques.
The cathode will be fabricated from high-purity oxygen-free copper particularly screened for low uranium and thorium contaminations. 
The field cage is designed to have only low-radioactivity acrylic facing the active volume to avoid background events from the PCB and resistors.
Surface contaminations of the Micromegas are more challenging to measure and control.
In this study, we used microbulk Micromegas~\cite{MM}, which is created from Kapton and copper using the PCB lithography technique and has been proven to have good performance and low surface radioactivity\cite{MMbackground}.
Furthermore, the surfaces of detector components, particularly those facing the active volume of the TPC, will be thoroughly cleaned to reduce the emission of $\alpha$ particles.
The inner volume of the Screener3D will be always filled or flushed with argon gas to avoid radon contamination from the air. 
In our sensitivity study later, we assume all the detector components are created from low-radioactivity materials and no secondary surface contaminations are introduced in construction.

\begin{figure}[tb]
	%\vspace{-0.01cm}
	\centering
	\includegraphics[width=\columnwidth]{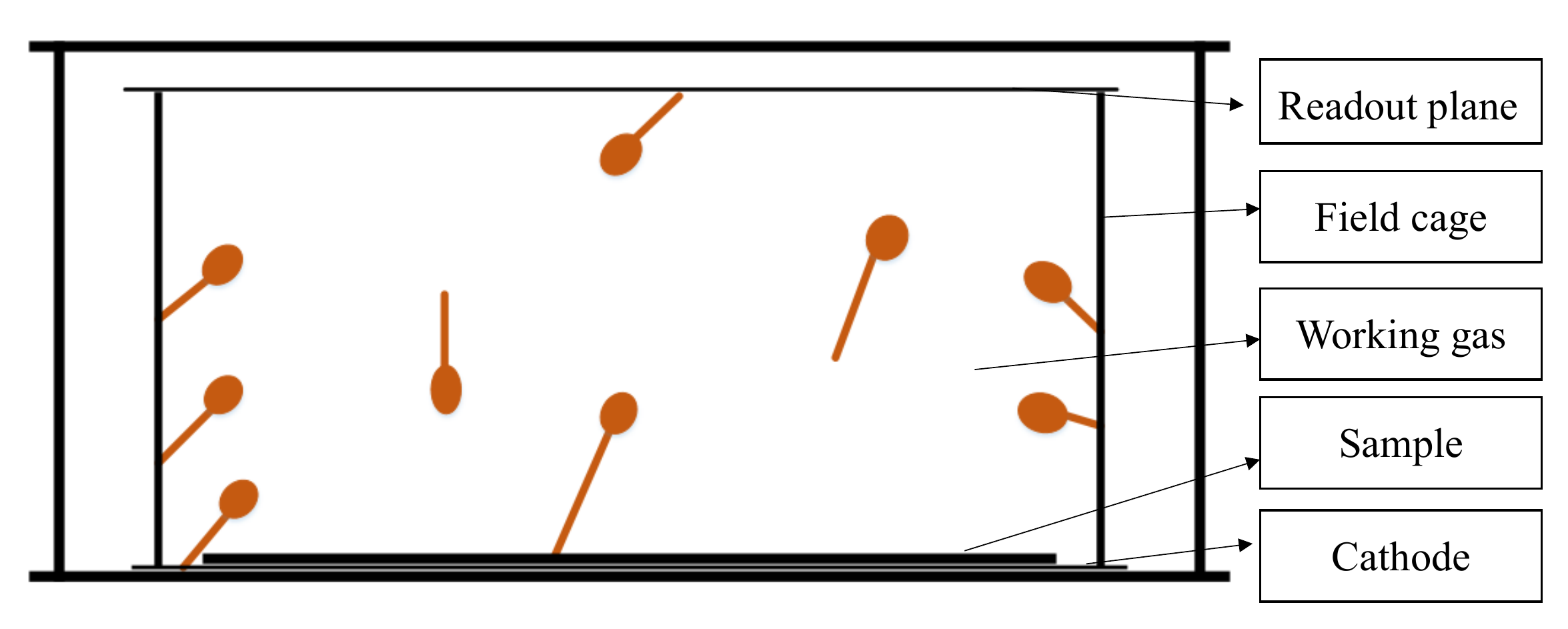}
	\caption{Illustration of $\alpha$ particle trajectories in a gaseous TPC. $\alpha$ particles travel mostly in a straight line in 1 bar of argon and 5\% isobutane gas with a prominent Bragg peak at the end. We utilize this feature for background suppression in our detector.}
	\label{fig:TrackFeature}
\end{figure}

In addition to minimizing the radioactive emission from detector components, we can further suppress the background with topological information from particle tracks, which is the key feature of Screener3D.
Tracks of $\alpha$ particles from a sample and various TPC components are shown in Fig.~\ref{fig:TrackFeature}.
The high energy $\alpha$ tracks are mostly straight in 1 bar of argon.
A large blob at one end of the track highlights the $dE/dx$ increase when $\alpha$ particles stop.
$\alpha$ events originating from the sample would have starting points from the sample surface and an upward-going direction in our detector geometry.
For the contaminations of the field cage, $\alpha$ tracks enter the active volume from the side and may travel upward or downward.
Surface $\alpha$ events from the readout plane point downward and can be easily distinguished from those of the samples. 
During sample measurement, the majority of the surface area of the cathode is covered by a sample and no $\alpha$ particle from the cathode could enter the active volume excepted the very edge (Fig.~\ref{fig:TrackFeature}).
With the distinguished track features, we can significantly suppress $\alpha$ backgrounds from the detector and achieve a higher sensitivity. 
In the following section, we demonstrate the suppression power using simulation data.

\section{Expected screening sensitivity}\label{sec:sensitivity}
The screening sensitivity of Screener3D depends on the background event rates recorded by the detector. 
In this section, we describe the setting up of  a detector model in the simulation framework Geant4~\cite{Geant4} and major background sources for $\alpha$ particle measurement.
The energy deposition process and gaseous detector response were implemented in the simulation. 
Signal selection and background suppression efficiencies were first calculated using energy cuts and particle identification.
The unique tracking capability of the gaseous detector was then exploited to distinguish $\alpha$ particles from different origins and with different orientations to further reduce the $\alpha$ background rate.
We finally present the material screening sensitivity for surface contamination of $\alpha$-emitting impurities.

\subsection{Screener3D background}
\begin{figure}[tb]
	%\vspace{-0.01cm}
	\centering
	\includegraphics[width=\columnwidth]{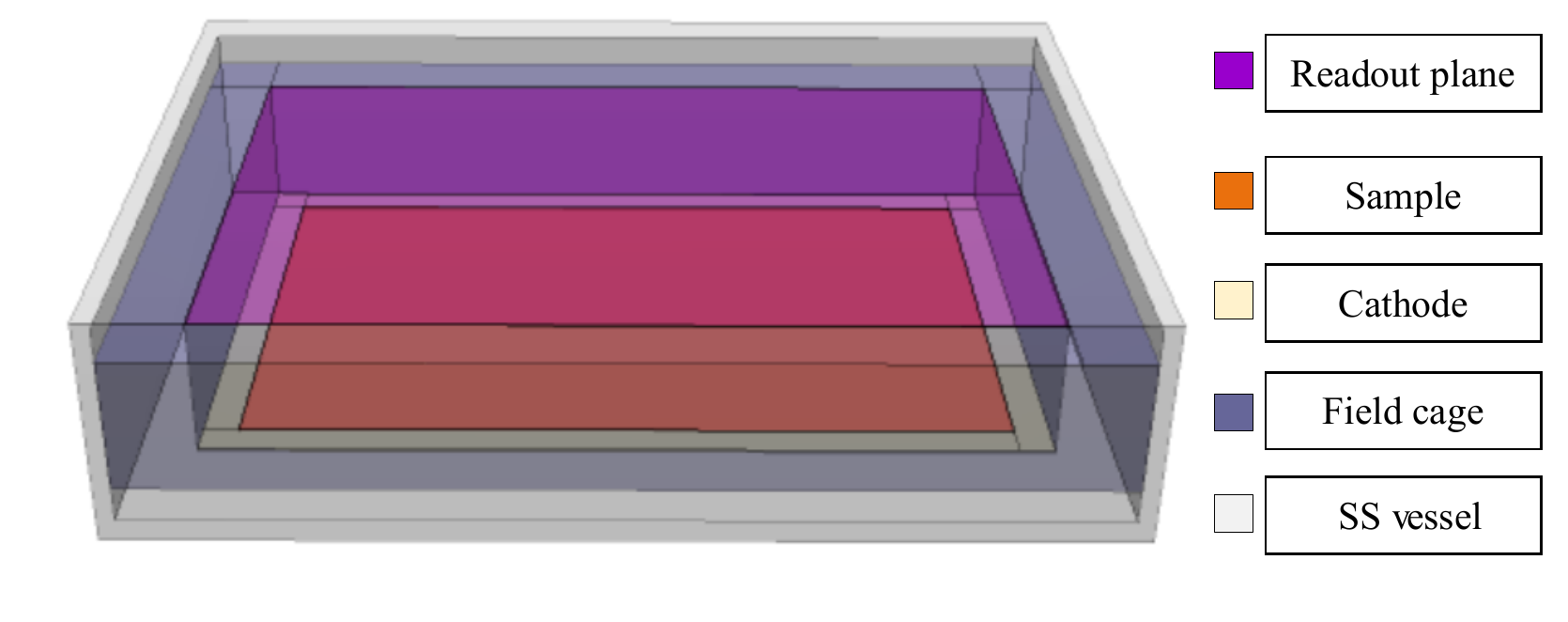}
	\caption{Geometry  of main components of Screener3D as constructed in the Geant4 simulation. All the dimensions followed the schematic design as shown in Fig.~\ref{fig:DrawingTPC}.}
	\label{simulation_geometry}
\end{figure}

\begin{table*}[tb]
	\def\arraystretch{1.5} 
	\tabcolsep 5pt 
	\caption{Radioactivity level of bulk material in Screener3D }
	\label{tab:radioactivities}
	\begin{tabular}{lccccccc}
		\toprule
	     Material             &\multicolumn{7}{l}{Radioactivity} \\
	     \cline{2-8}
	                          
	                          &$^{238}$U     &$^{232}$Th    &$^{40}$K        &$^{60}$Co       &$^{137}$Cs       &$^{222}$Rn          &$^{39}$Ar           \\
	                          &(mBq$\cdot$kg$^{-1}$) &(mBq$\cdot$kg$^{-1}$)  &(mBq$\cdot$kg$^{-1}$) &(mBq$\cdot$kg$^{-1}$) &(mBq$\cdot$kg$^{-1}$) &($\mu$Bq$\cdot$m$^{-3}$)  &(mBq$\cdot$kg$^{-1}$) \\
	     \hline
	     Gaseous argon        &0.0018         &0.0004        &               &                 &                  &10                  &10$^{3}$\\
	     Acrylic              &0.003          &0.003         &               &                 &                  &                    &\\
	     Oxygen-free copper   &0.38           &0.51          &4.00           &0.20             &0.16              &                    &\\
	     Stainless            &1.70           &2.74          &13.95          &1.03             &2.36              &                    &\\
		\bottomrule
	\end{tabular}
\end{table*}

Fig.~\ref{simulation_geometry} shows the detector geometry constructed in the Geant4 simulation~\cite{Geant4}. 
The setup consisted of the sensitive volume, field cage, readout plane, cathode, and SS vessel with dimensions identical to the detector's conceptual design.
The sensitive volume in the center was a $60\times40\times10$~cm$^3$ cuboid with 1 bar argon and 5\% isobutane mixture.
The readout plane is expected to be instrumented using microbulk Micromegas, which has complicated mechanical structures, such as a lithographed copper mesh, avalanche holes, and underlining PCB patterns~\cite{MM}. 
The structure was substituted with a 0.1~mm-thick copper. 
The readout plane on the top of the sensitive volume consisted of 6 Micromegas modules and was 58.0~cm long and 38.5~cm wide.
The cathode, created from 2~mm-thick oxygen-free copper, had the same surface area and faced the readout plane beneath the sensitive volume.
A 4~cm-thick acrylic frame surrounding the sensitive volume from four sides represented the mechanical structure of the field cage and acrylic frame.
We placed a 55$\times$35$\times$0.01~cm$^{3}$ sample on the cathode in the simulation.
All the components described above were enclosed in a 1~cm-thick SS vessel with an inner volume of $80\times 60 \times 15$~cm$^3$.

We simulated major background contributions from surface and bulk containments of all components.
The energy deposition process of $\alpha$, $\beta$, and $\gamma$ particles in the TPC were simulated in Geant4.
For the study of $\alpha$ contamination on the sample surface, we expected the $\beta$ and $\gamma$ background events from detector bulk material to have a marginal impact.
We confirmed this in our simulation first before focusing on the $\alpha$ background from surfaces of detector components.

Major bulk background sources such as $^{238}$U, $^{232}$Th, and other radio-contaminations in different materials are listed in Table~\ref{tab:radioactivities}.
The radio-contamination of oxygen-free copper and SS followed measurements of the PandaX Collaboration~\cite{Zhang:2018xdp}.
For acrylic, the radioactivities were assumed to be the same as that used by Reference~\cite{Juno_acrylic}.
The working gas medium for $\alpha$ contamination measurement was argon.
The radio-purity of commercial argon gas varies significantly from different vendors or even different batches of the same vendor. 
For $^{238}$U and $^{232}$Th in argon, we adopted values from GERDA~\cite{BUDJAS2009755}.
Moreover, atmospheric argon contains approximately 1~Bq$\cdot$kg$^{-1}$ of cosmogenic $^{39}$Ar~\cite{Agnes:2015ftt}, which is a concern for $\beta$ measurement but will not introduce background in the high-energy region for $\alpha$ measurement.

$^{222}$Rn in argon is listed separately from the parent $^{238}$U chain.
Radon is a noble gas and is frequently at the $\mu$Bq-mBq$\cdot$m$^{-3}$ level in commercial argon gas supply.
The contamination can be reduced by at least three orders of magnitude after passing through an activated carbon cold trap~\cite{Simgen}.
We assumed a concentration of $^{222}$Rn of 10~$\mu$Bq$\cdot$m$^{-3}$ in our simulations.

\begin{figure}[tb]
	%\vspace{-0.01cm}
	\centering
	\includegraphics[width=\columnwidth]{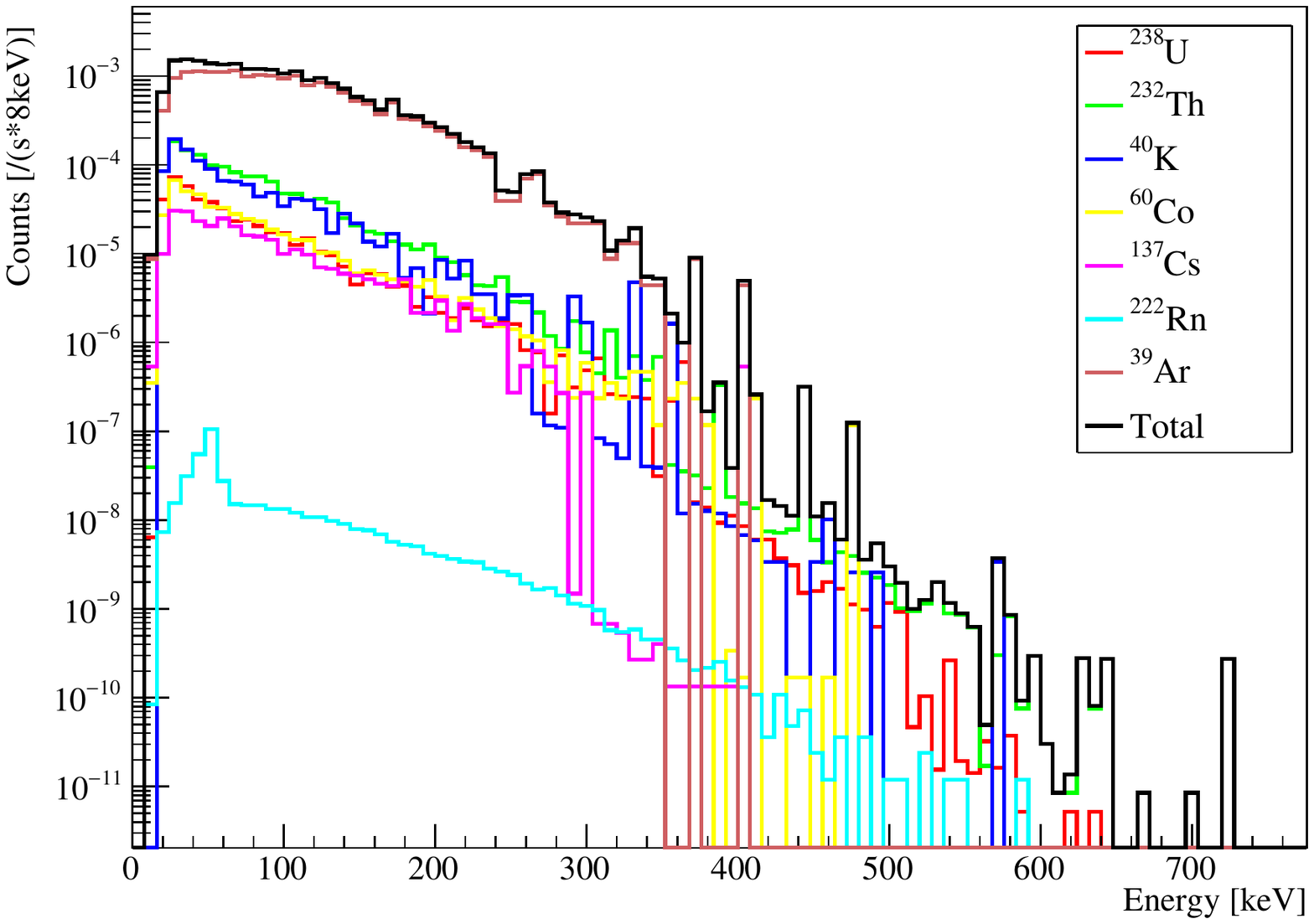}
	\caption{Low-energy spectra of Screener3D. Main contributions in the energy range were from $\gamma$ and $\beta$ events.}
	\label{fig:g4spec_gamma_e}
\end{figure}

Fig.~\ref{fig:g4spec_gamma_e} shows the energy deposition of $\beta$ and $\gamma$ events in Screener3D.
Owing to the relatively small $dE/dx$ of $\beta$ and $\gamma$ particles, the typical energy deposition is less than 1~MeV, which is distinctively different from that of $\alpha$ particles. 
Therefore, we defined the region of interest (ROI) for $\alpha$ particle measurement as above 1~MeV.
The upper limit for $\alpha$ ROI was 10~MeV, which is above the highest $\alpha$ energy of naturally occurring radioactivity of the decay chains of uranium and thorium.

Only the $\alpha$-emitting surface directly facing the sensitive volume, including argon gas, readout plane, field cage, and cathode contribute to the $\alpha$ background.
$\alpha$ particles with energy below 10~MeV travel less than 20~$\mu$m in copper and less than 100~$\mu$m in acrylic.
Therefore, $\alpha$ contamination near the surface were our main focus.
The concentration of $\alpha$-emitting radio-isotopes, particularly decay chains of $^{238}$U and $^{232}$Th, on the surface is frequently higher than that in the bulk, owing to secondary contamination. 
The surface radioactivities of acrylic, oxygen-free copper, and Micromegas readout plane are shown in Table~\ref{tab:Sur_alpha_radioactivities}.
We deduced a conservative surface contamination level of acrylic by comparing two sets of acrylic contamination levels with and without the surface cleaning
from the JUNO Collaboration~\cite{Juno_acrylic}.
For copper, we adopted the maximum values measured by the CUORE Collaboration in dedicated bolometer arrays~\cite{CUORE_copper}.
The radioactivity levels of the readout plane followed the results of microbulk Micromegas measured using the BiPo-3 detector~\cite{MMbackground, PandaX-III}.

The depth profile of surface contaminations can be approximated using an exponential curve, 
\begin{equation}
\rho={\rho_{0}e^{-\frac{x}{\lambda}}}+b,
\label{equ:impdis}
\end{equation}
where $\rho$$_{0}$ is the surface contamination per unit volume at the very surface, $x$ is the distance into the bulk material, $\lambda$ is the characteristic penetration depth of the contamination, and $b$ is the contamination level in the bulk material.
At the very surface, the contamination level is $\rho_0+b$~\cite{CUORE_copper}.
$\lambda$ may vary significantly with the contamination type and causes of contamination.
The depth profile can be verified and the exact value of $\lambda$ determined by measuring the peak shape of $\alpha$ spectrum, but they are challenging experimentally.
In our simulation, we set $\lambda=0.1\,\mu$m, which is a shallow distribution, but the effects can still be experimentally identified.
Additionally, we compared our results with $\lambda=1\,\mu$m, to study the influence of the penetration depth of the contamination.
For both $\lambda$ values, the total amount of the surface contaminations in the materials was the same.

\begin{table}[tb]
	\def\arraystretch{1.5} 
	\tabcolsep 7pt 
	\caption{Radioactivity level of surface in Screener3D}
	\label{tab:Sur_alpha_radioactivities}
	\begin{tabular}{lcc}
		\toprule
		Material            &\multicolumn{2}{c}{Radioactivity(mBq$\cdot$m$^{-2}$)} \\
		                    \cline{2-3}
		                    &$^{238}$U    &$^{232}$Th  \\
		\hline
        Acrylic	            &0.25         & 0.08  \\
		Oxygen-free copper  &1.30         & 1.30 \\
		Readout plane       &0.45         & 0.14 \\
		\bottomrule 
	\end{tabular}
\end{table}

\begin{figure}[tb]
	\centering
	\subfigure[Geant4-simulated energy spectra assuming $\lambda=0.1\,\mu$m]{
		\includegraphics[width=0.9\columnwidth]{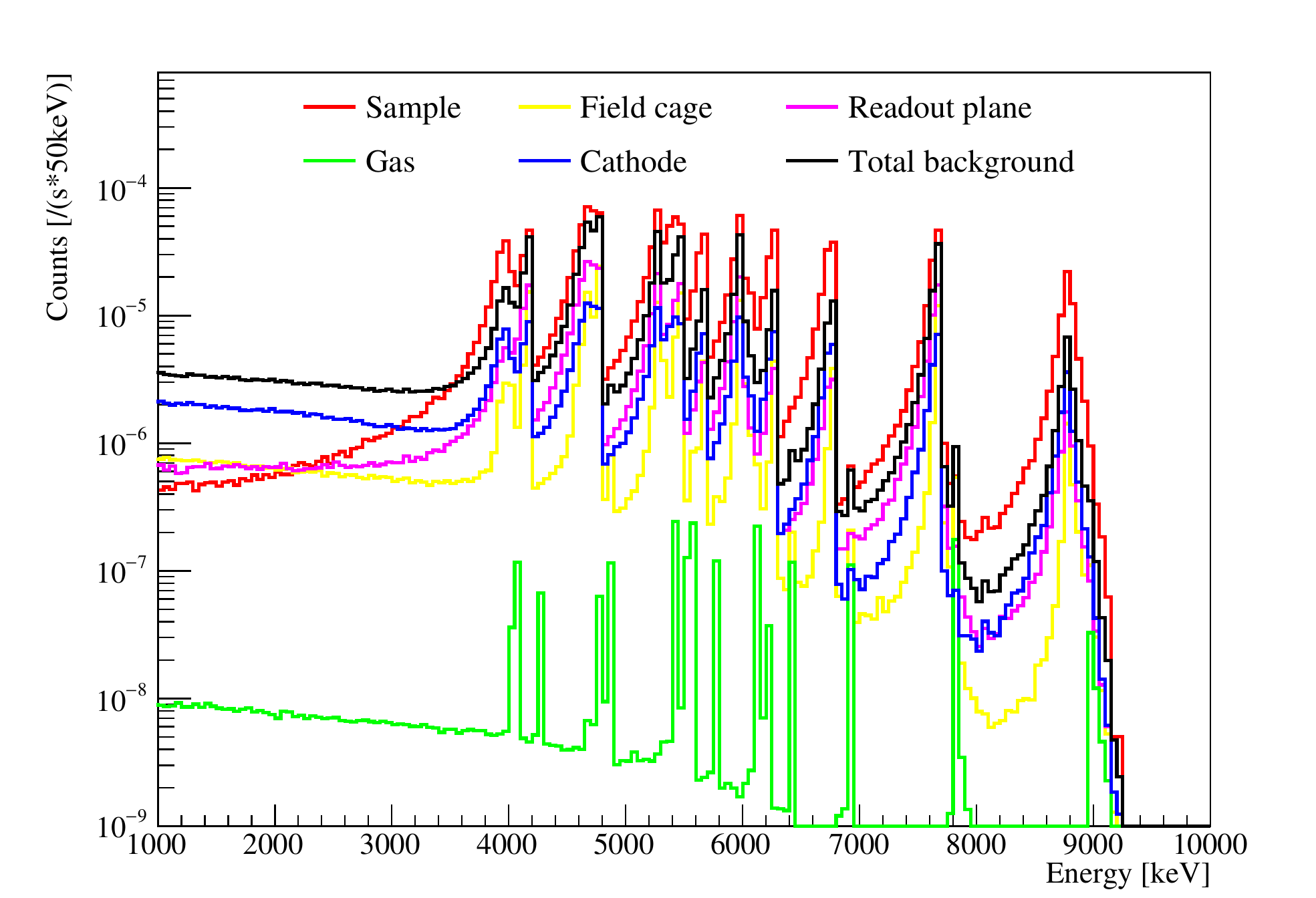}} 
	\subfigure[Geant4-simulated energy spectra assuming $\lambda=1\,\mu$m]{
		\includegraphics[width=0.9\columnwidth]{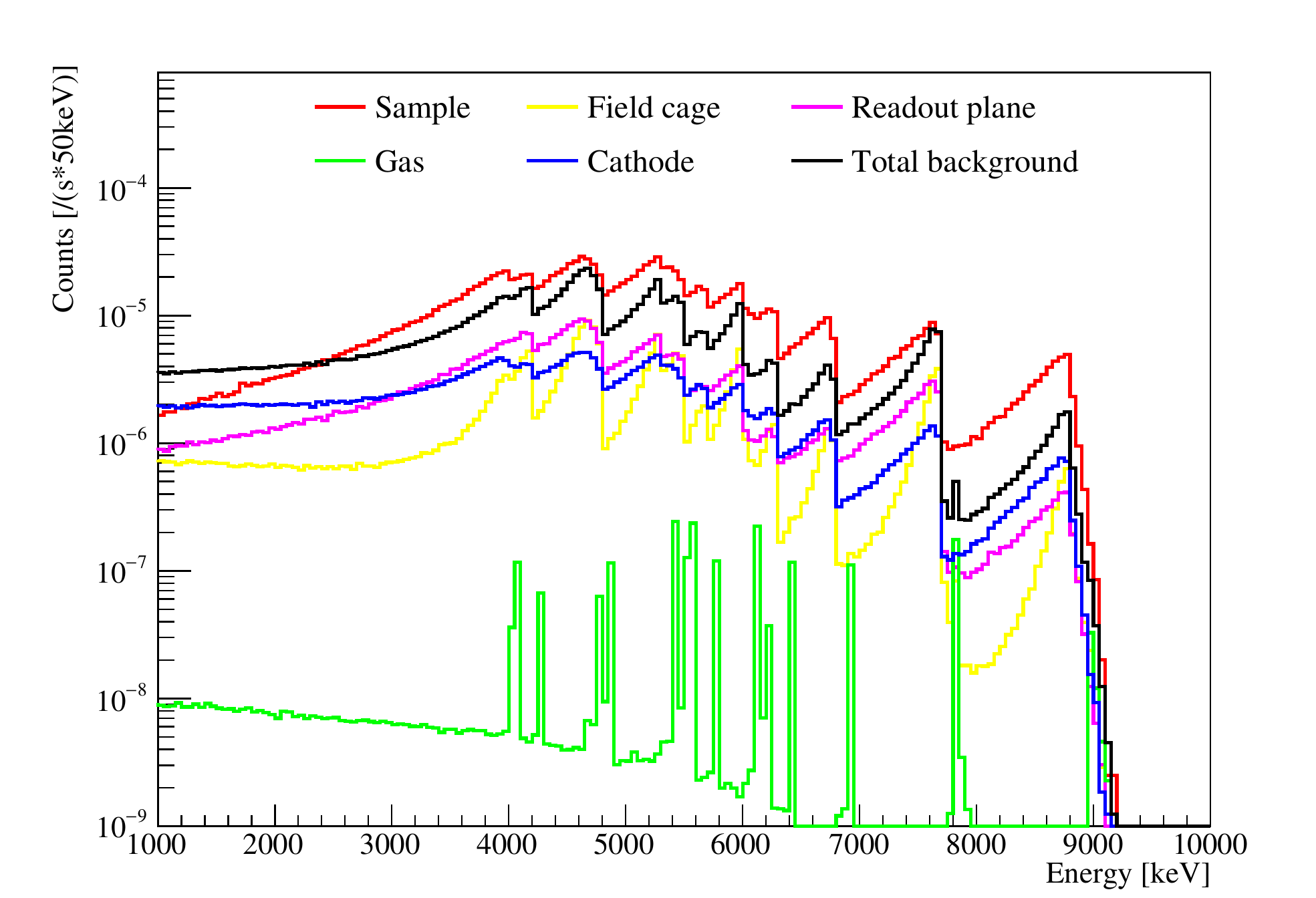}} 
	\caption{Geant4-simulated energy spectra of Screener3D. For a longer penetration depth of $\lambda = 1\,\mu$m, $\alpha$ particles are more likely to lose partial energy before escaping the surface.
		Therefore, the tail of the characteristic peak is more apparent in figure (b).}
	\label{fig:g4spec_alpha}
\end{figure}

Fig.~\ref{fig:g4spec_alpha} shows the energy spectra of a copper sample on the cathode plane and different detector components for both $\lambda$ values.
The count rates of the sample, field cage, readout plane, and cathode were scaled with the surface area and the corresponding surface radioactivity level listed in Table~\ref{tab:Sur_alpha_radioactivities}.
For the gas medium, the spectrum was determined by the contamination level of radon and total argon mass.
The bulk radioactivity level used in the simulation is listed in Table~\ref{tab:radioactivities}.

In the ROI between 1 and 10~MeV, the total expected background rate was 82.4 (76.1) counts per day for $\lambda=0.1\,\mu$m ($1\,\mu$m).
Approximately 46.0\% (45.4\%) of the background was from the readout plane, 36.1\% (35.8\%) from the cathode, 17.6\% (18.5\%) from the field cage, and 0.3\% (0.3\%) from the working gas.
The difference in $\alpha$ peak shapes of Fig.~\ref{fig:g4spec_alpha} (a) and Fig.~\ref{fig:g4spec_alpha} (b) highlight the impact of depth profile.
For relatively radio-active samples with high contamination levels, we can potentially have a quantitative estimation of the depth profile.
In the two figures, the spectra from argon gas are identical.

\subsection{Detector response simulation}
\begin{figure}[tb]
	\centering 
	\subfigure[Simulated waveforms recorded by Micromegas]{
		\includegraphics[width=\columnwidth]{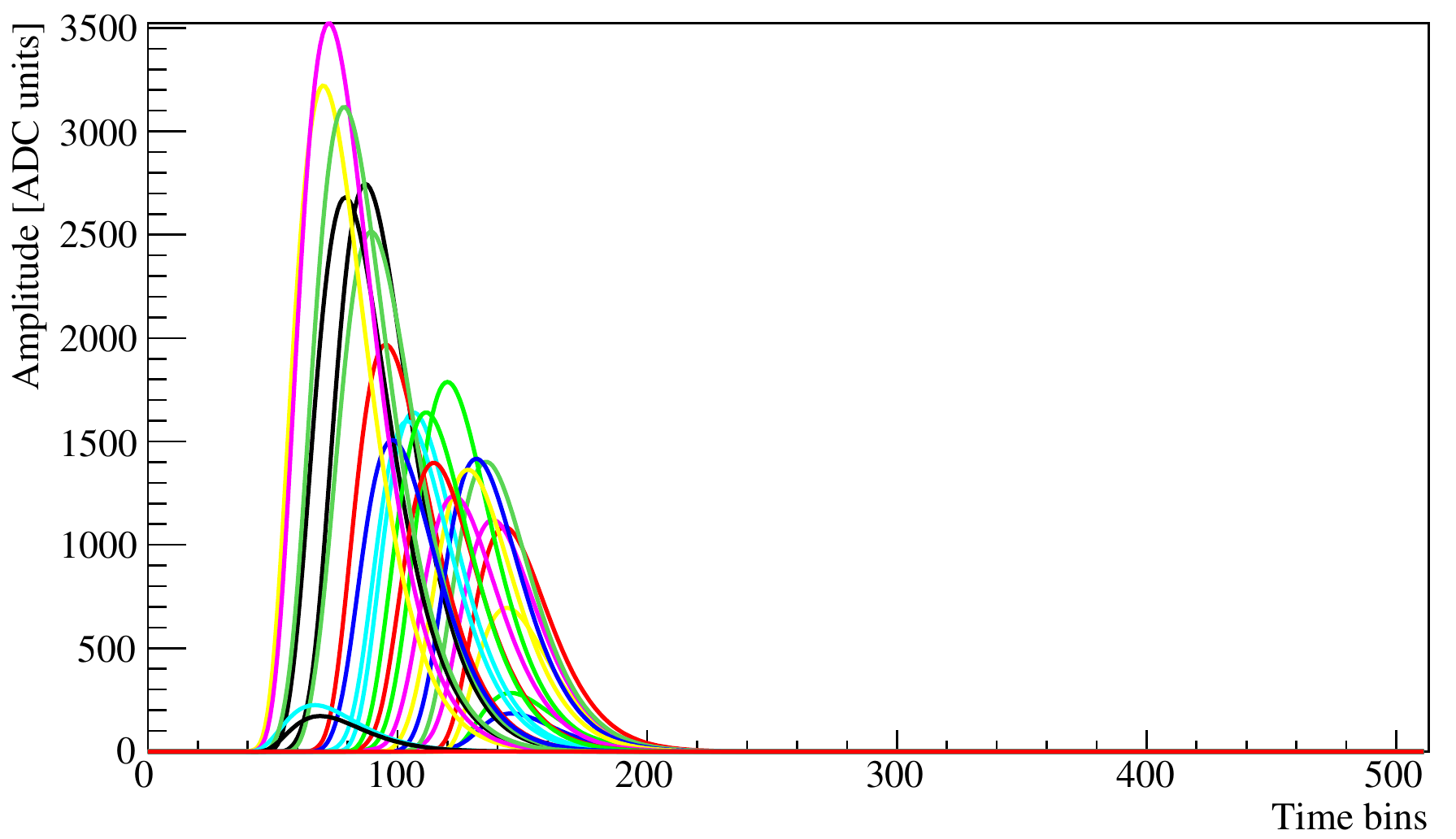}} 
	\subfigure[Waveforms of an $\alpha$ event in a TPC measured with a microbulk Micromegas readout plane.]{
		\includegraphics[width=\columnwidth]{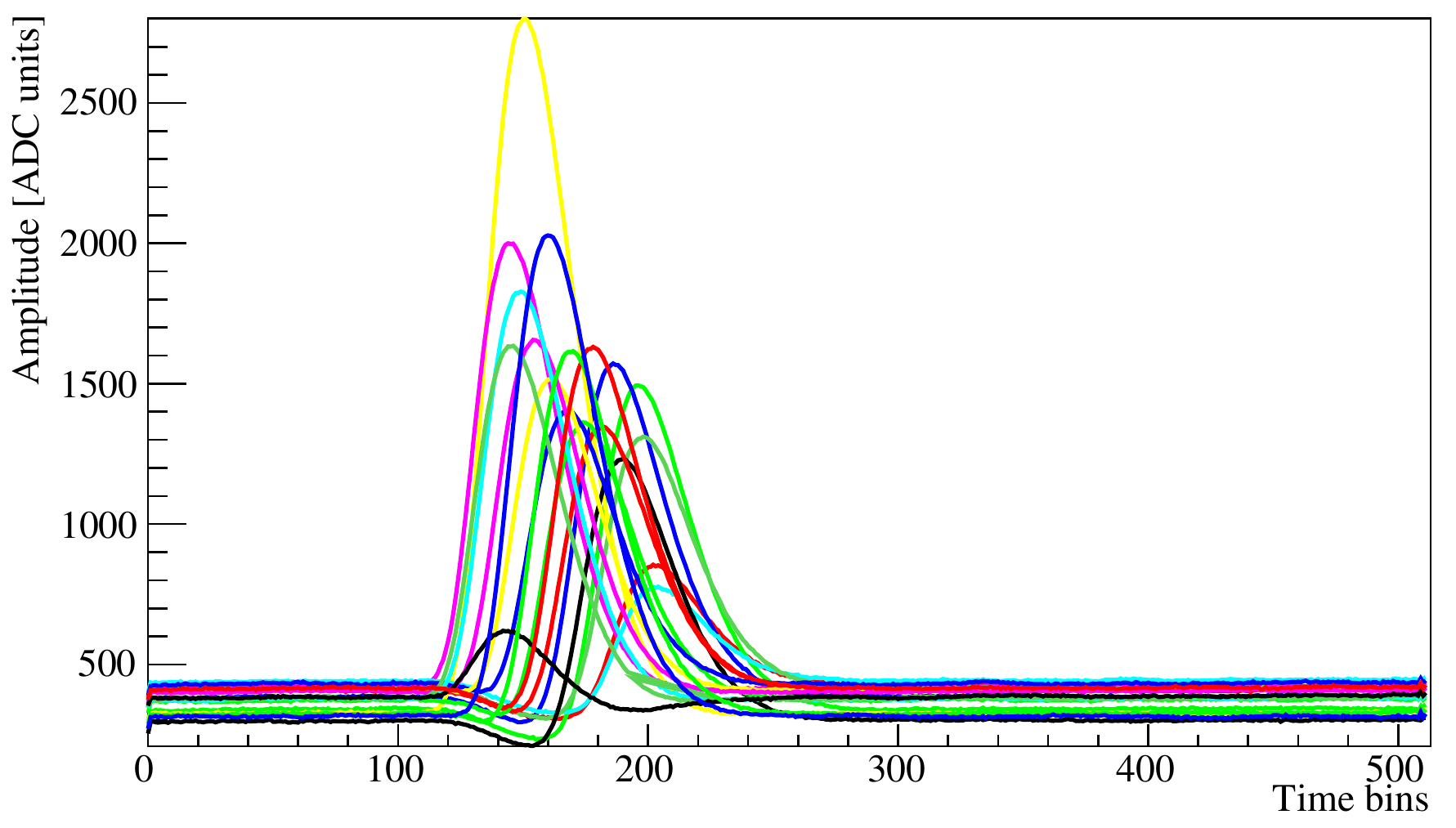}} 
	\caption{Simulated (a) and measured (b) waveforms of an $\alpha$ event. 
		The measurement was performed with 1 bar of argon and 5\% isobutane gas mixture.}
	\label{fig:simsExpPulse} 
\end{figure}

Detector response, including electron diffusion, energy resolution, readout scheme, and electronics response, was added to the Geant4 simulation data to produce mock detector data for downstream analysis.
The detector response was simulated using the REST framework~\cite{REST}, which was developed and used by PandaX-III~\cite{PandaX-III} and other gas TPC projects.
If we use an $\alpha$ particle traveling in gas as an example, Geant4 recorded the differential energy deposition $dE/dx$ along the trajectory. 
In REST, the differential energy deposition was converted to the number of ionization electrons in gas.
The ionization electrons diffused while drifting to the readout plane. 
REST interfaced with the Garfield program~\cite{Garfield} to calculate diffusion coefficients.
For argon/isobutane(5\%) mixture at 1 bar and under an electric field of 100~V/cm, the transverse (longitudinal) diffusion coefficient was 0.048 (0.044)~$\sqrt{\text{cm}}$. 
The diffusion of the electrons was then calculated as $D_{ic}\sqrt{L_{d}}$, where $D_{ic}$ is the diffusion coefficient and $L_{d}$ is the drifting distance.
When the diffused electrons reached the readout plane, the position and arrival time were recorded. 
Following the strip-readout schemes of Micromegas in the actual detectors, ionization electrons were grouped by strips, and pulses were generated strip by strip (Fig.~\ref{fig:simsExpPulse} (a)).
The colored pulses represent signals from different strips.
The pulse amplitude in Fig.~\ref{fig:simsExpPulse} is proportional to the number of electrons collected per strip.
The amplitudes were smeared according to a Gaussian function to account for an energy resolution of 3\% full width at half maximum (FWHM) at 2.5~MeV in the detector response simulation.
Pulse widths were determined from electronics shaping and a 1~$\mu$s shaping time was used (Fig.~\ref{fig:simsExpPulse}~(a)).
The relative timing among pulses indicated the arrival time of electrons.
Under a relative constant drift velocity of approximately 30 mm/$\mu$s, the arrival time denoted the relative energy deposition position in the drift direction.
For a sampling rate of 50~MHz and a record length of 512 sample points per signal, we recorded approximately 10~$\mu$s of data per signal window.
For a drift length of 10~cm in our TPC, the signal window was more than adequate, as shown in the figure.
We also implemented triggers in REST, but the trigger threshold effect was negligible for $\alpha$ events in the ROI.

For comparison, Fig.~\ref{fig:simsExpPulse}~(b) shows an $\alpha$ event recorded by a gaseous detector.
The detector was equipped with a $20\times20$~cm$^{2}$ Micromegas and filled with 1 bar of argon and 5\% isobutane gas mixture.
The height of the drift volume was 10~cm, the same as in our simulation.
Our mock pulses re-produced the key features of detector pulses.
Next, the mock pulses of each simulated event were analyzed using the same procedure as detector data. 

\begin{figure}[tb]
	%\vspace{-0.01cm}
	\centering
	\subfigure[Simulated energy spectra with detector response assuming $\lambda=0.1\,\mu$m]{
		\includegraphics[width=\columnwidth]{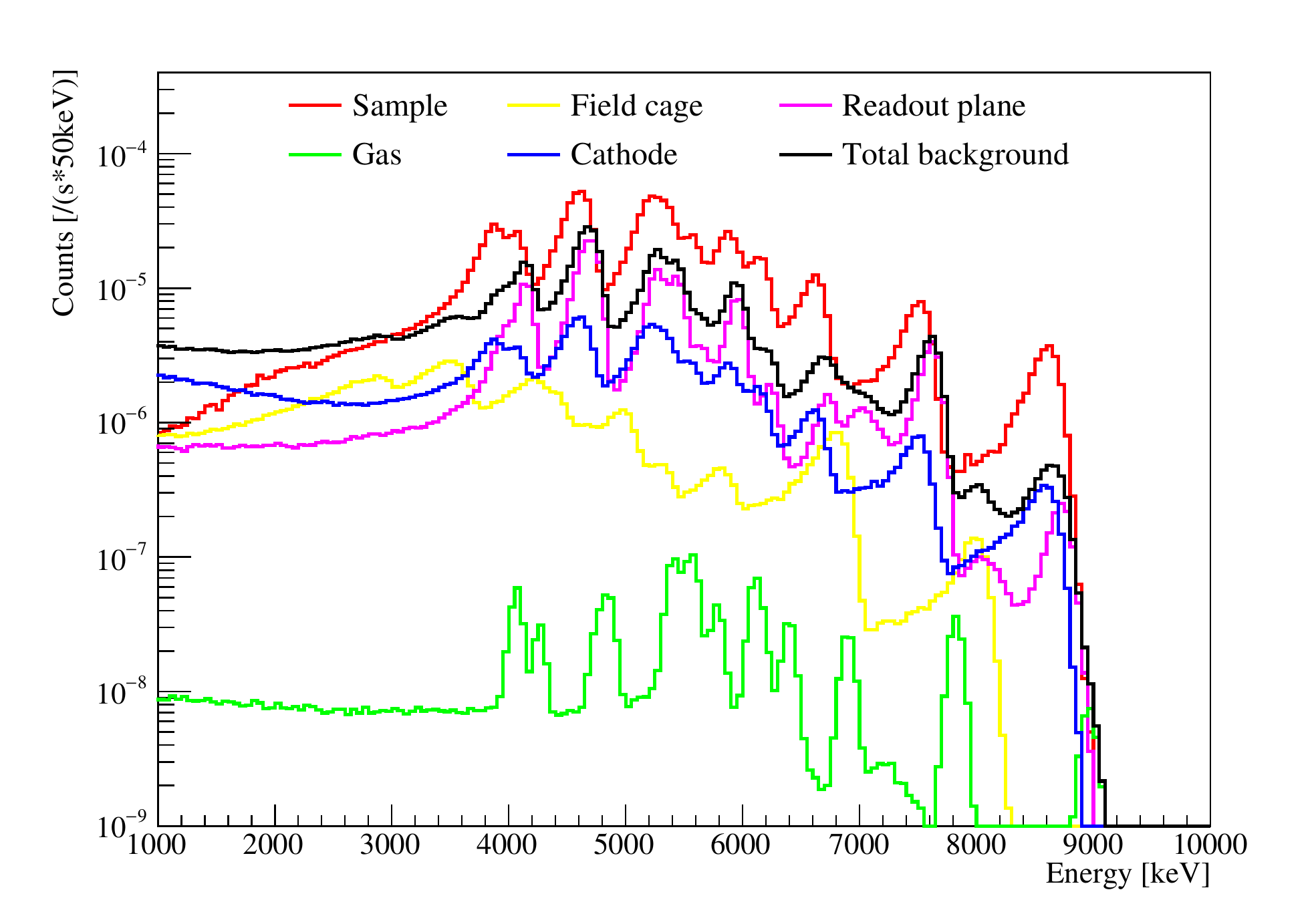}} 
	\subfigure[Simulated energy spectra with detector response assuming $\lambda=1\,\mu$m]{
		\includegraphics[width=\columnwidth]{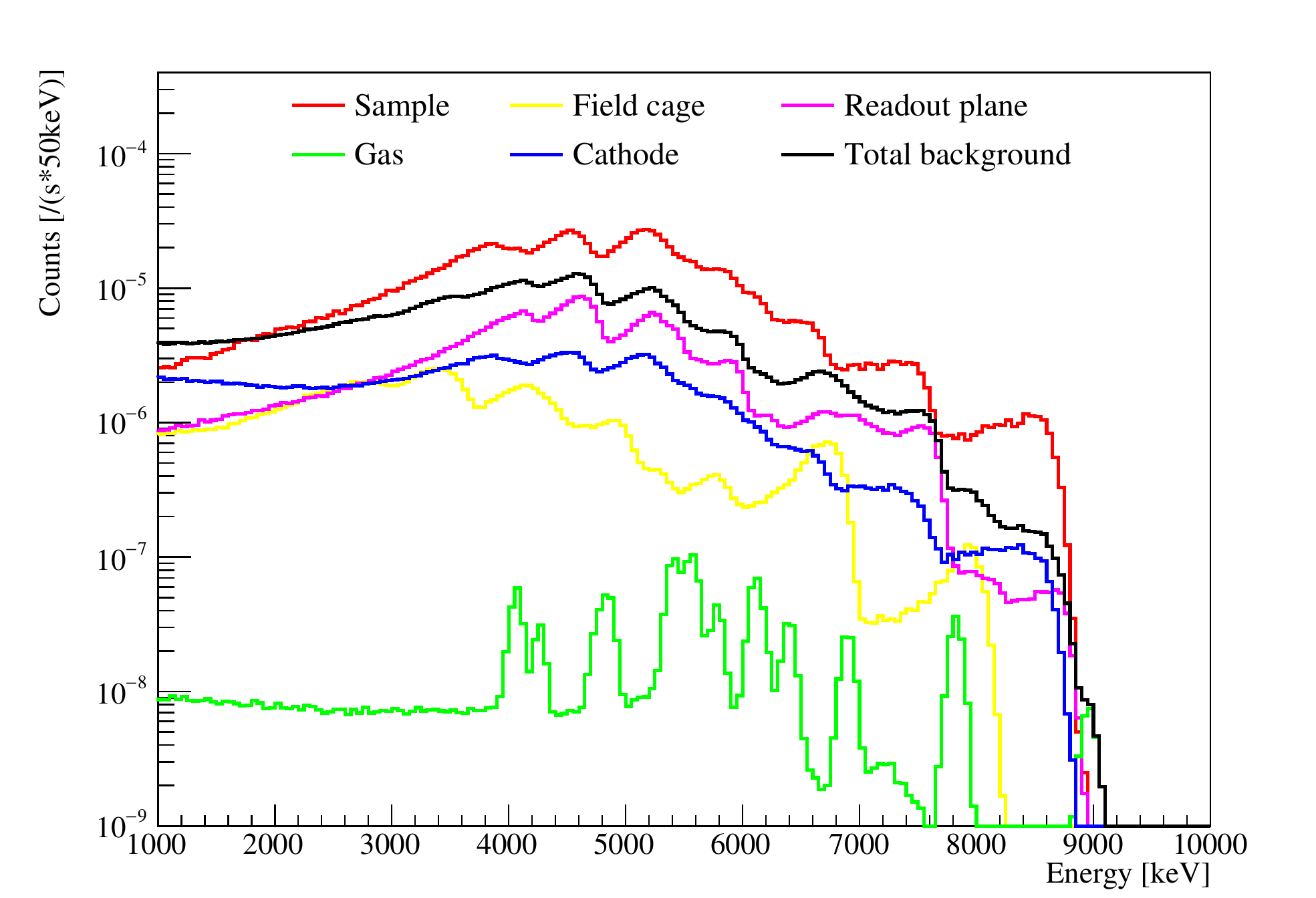}} 
	\caption{Simulated energy spectra of Screener3D with detector response added.}
	\label{fig:restspec}
\end{figure}

Energy was reconstructed using mock pulses and the resulting spectra are shown in Fig.~\ref{fig:restspec}.
Compared with  Fig.~\ref{fig:g4spec_alpha}, the $\alpha$ peaks were broadened owing to the detector resolution.
Since the effective area of the readout plane did not cover the full area of the field cage, partial energy of background events from the field cage were not be recorded. 
Therefore, the spectrum of background events from the field cage shifted to a lower energy compared with that in Fig.~\ref{fig:g4spec_alpha}.
Meanwhile, a small peak was observed in front of some characteristic peaks from the readout plane.

\subsection{Track reconstruction}
\begin{figure}[tb]
	\centering 
	\subfigure[Geant4-simulated tracks of an $\alpha$ particle]{
		\includegraphics[width=\columnwidth]{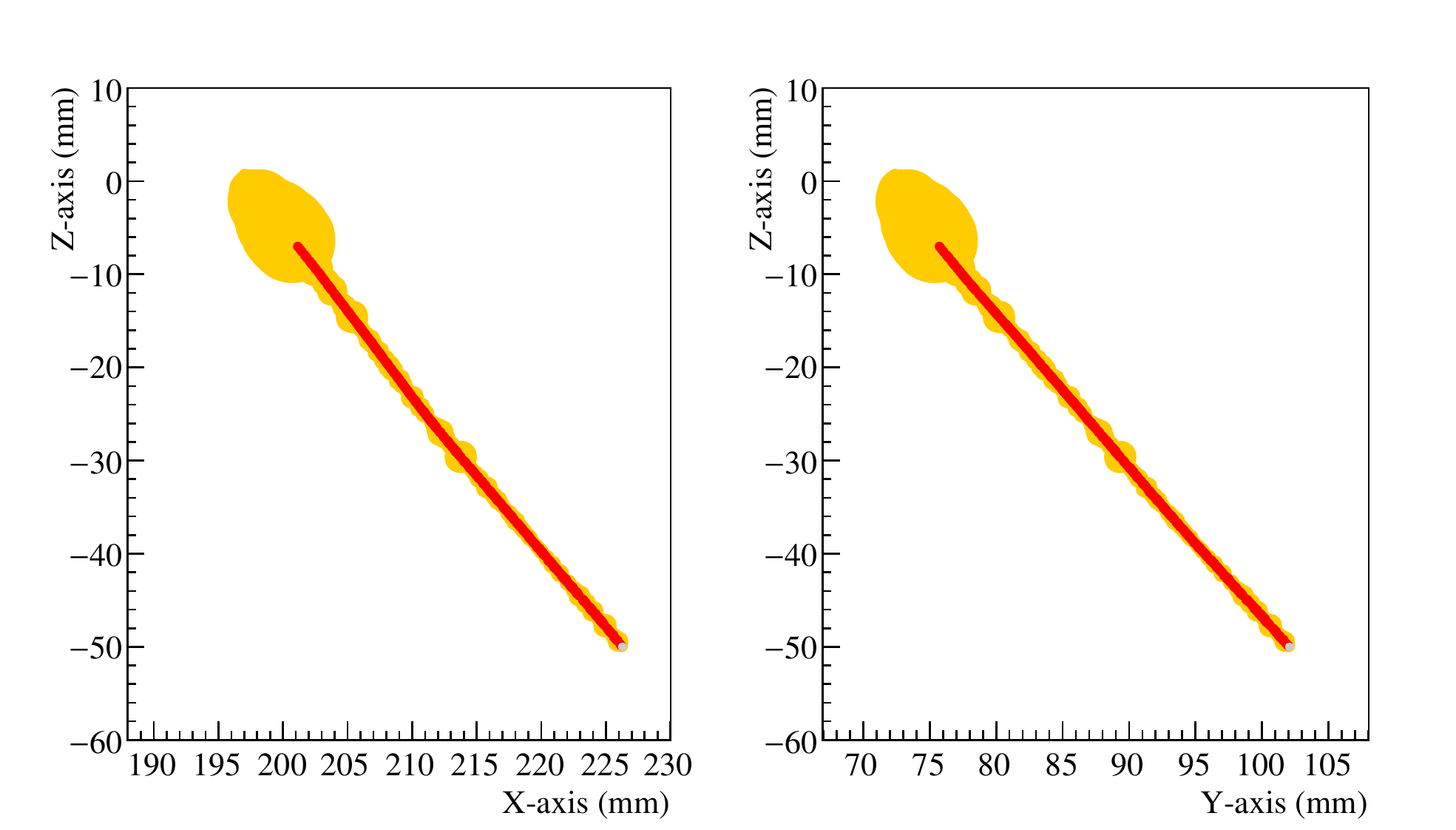}} 
	\subfigure[Hits recorded using Micromegas strips and the reconstructed track]{
		\includegraphics[width=\columnwidth]{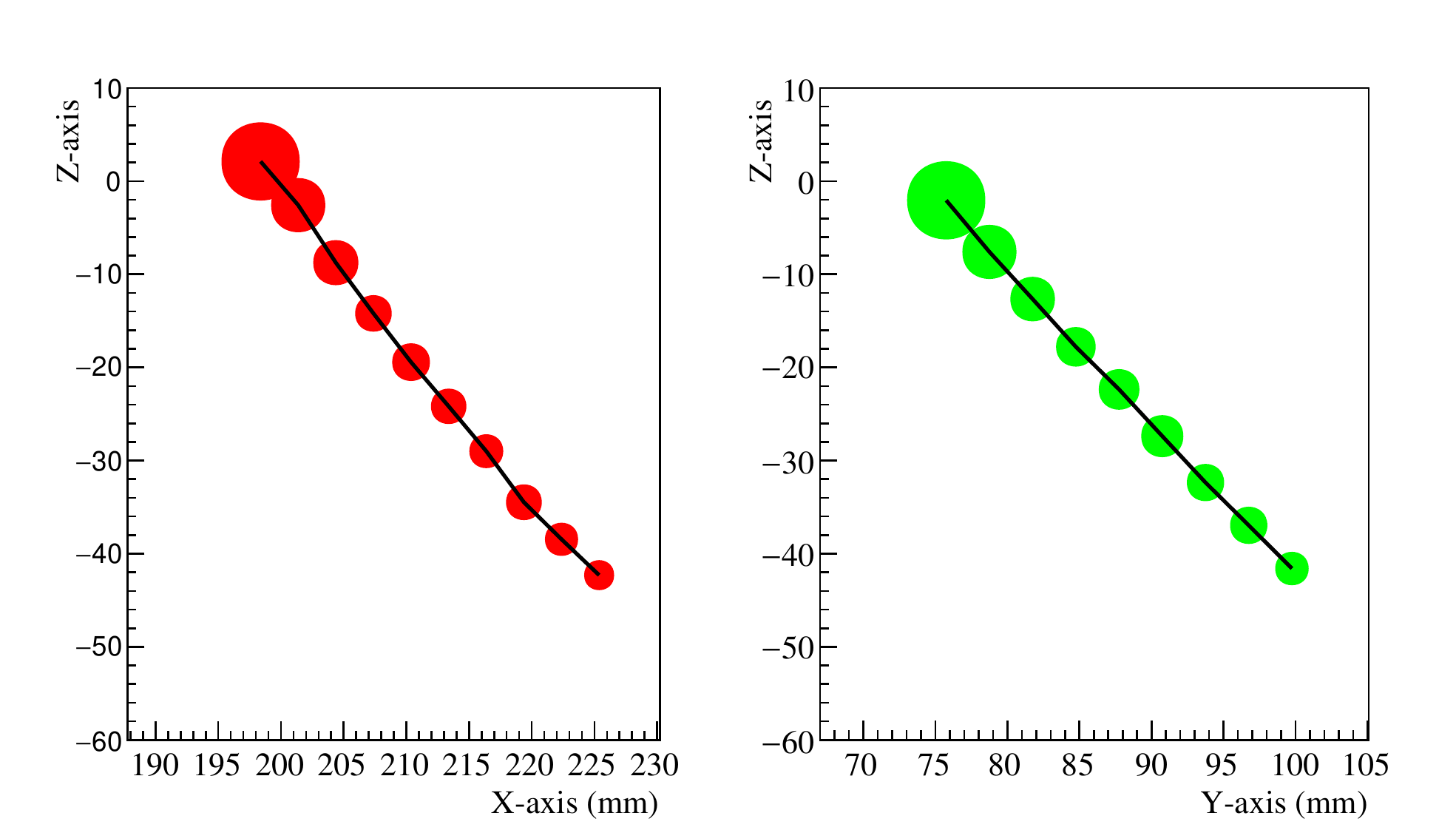}}
	\caption{Geant4-simulated (a) and reconstructed (b) tracks of an $\alpha$ particle. In (a), the straight red line represents the $\alpha$ particle track and yellow dots represent the relative size of energy deposition. In (b), each red/green dot represents a triggered strip signal and the size of the dot represents the amount of deposited energy in the strip. The black line is the reconstructed track by connecting nearby hits.
	}
	\label{tracks}
\end{figure}

The main application of mock pulses is track reconstruction.
As mentioned earlier, the amplitude and timing of pulses of different strips contain all the information we would collect from an actual detector.
Therefore, this information can be used to reconstruct the particle tracks and differential energy loss along the tracks in a TPC.
Track reconstruction was performed in the X--Z and Y--Z planes, where Z denotes the drift direction and X and Y represent transverse directions. 
In 1 bar argon, $\alpha$ particles would rarely be scattered by a large angle and the tracks are mostly straight.
Fig.~\ref{tracks} (a) shows an $\alpha$ track in the X--Z and Y--Z planes.
The red trace represents the trajectory and yellow dots represent the energy deposition vertex.
The size of yellow dots illustrates the relative amount of energy deposition.
The Bragg peak is prominent at the end of the trajectory on the top left of the figures.

In Fig.~\ref{tracks} (b), we show the reconstructed track in the X--Z and Y--Z planes of the same event as in Fig.~\ref{tracks} (a).
Each red (or green) point in the figures represents a triggered strip signal, and the size of the point represents the amount of deposited energy.
The X (Y) coordinate was determined from the position of the strip.
The Z-axis indicates the position of the ionized electron relative to that of the first electron to reach the readout plane, as determined using the drift velocity and pulse timing.
For long, near-straight $\alpha$ tracks, sorting by timing or X/Y position can reliably reconstruct the true tracks.
The black lines in Fig.~\ref{tracks} (b) represent a well-reconstructed track.
Track reconstruction can be challenging when the tracks are short. 
For example, the length of the X--Z and/or Y--Z tracks may be very short because of the projection angle.
The tracks may also be short when $\alpha$ particles only deposit partial energy in the active volume.

\subsection{Screener3D background suppression}

In our ROI from 1 to 10 MeV for $\alpha$ measurement, effectively all the $\beta/\gamma$ background were removed.
The ROI cut also removed 11\% of $\alpha$ background and maintained 98.9\% of the signal events.
Alpha background could be further suppressed using fiducial, angle, and hit-number cuts.
We explain the cuts in detail as follows.

We set the cut values to maximize the detector sensitivity $S_{d}$.
For measurements with a large number of backgrounds $B$, background fluctuation follows a Poisson distribution and equals $\sqrt{B}$.
Therefore, the detector sensitivity is proportional to $\epsilon_{s}/\sqrt{\epsilon_{b}}$, where $\epsilon_{s}$ is the signal detecting efficiency and $\epsilon_{b}$ is the efficiency in which background events are maintained through selection cuts.
For each cut, we selected the cut values that maximized the $\epsilon_{s}/\sqrt{\epsilon_{b}}$.

\begin{figure}[tb]
	%\vspace{-0.01cm}
	\centering
	\includegraphics[width=\columnwidth]{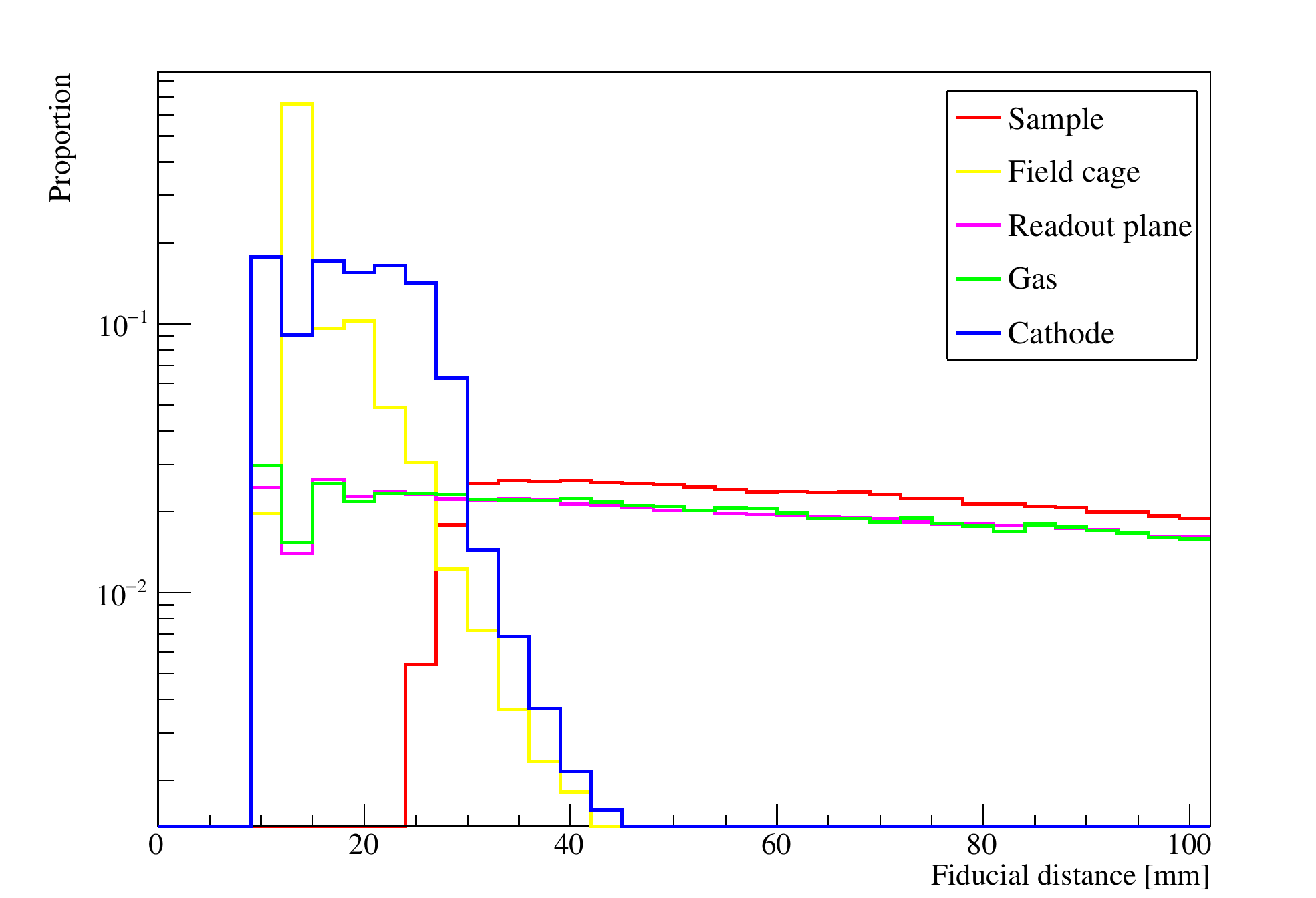}
	\caption{Distribution of tracks starting point. We created a cut at the fiducial distance of 27~mm to effectively remove the background events from the field cage and cathode.}
	\label{Fiducial}
\end{figure}

\begin{figure}[tb]
	\includegraphics[width=\columnwidth]{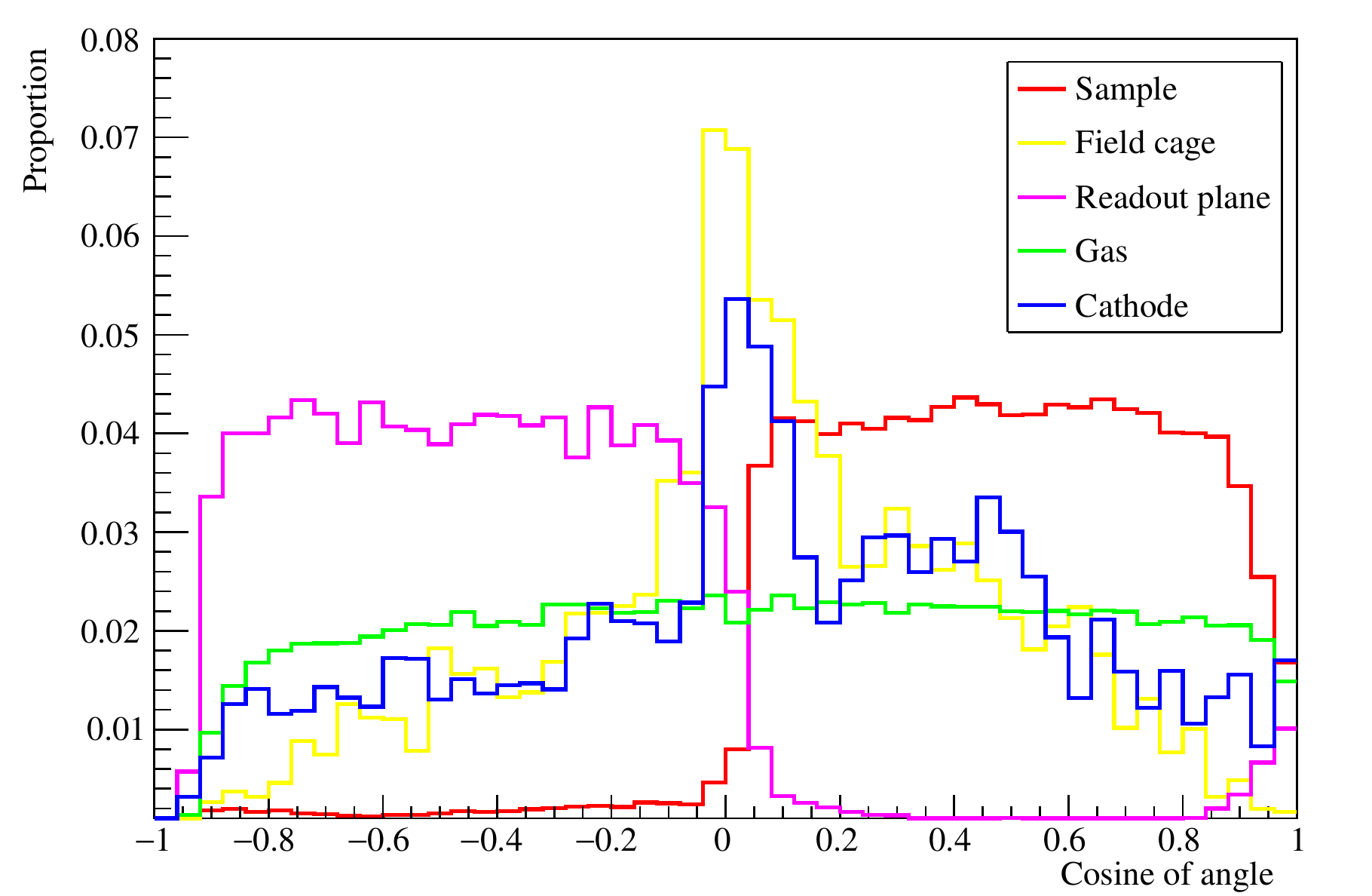}
	\caption{Distribution of track direction, defined as the cosine of the angle between the track and vertical direction. The distribution is for events that survived the fiducial cut.
		We created a cut at 0 to effectively remove the background events from the readout plane and gas.}
	\label{Angle}
\end{figure}

\begin{figure}[tb]
	%\vspace{-0.01cm}
	\centering
	\includegraphics[width=\columnwidth]{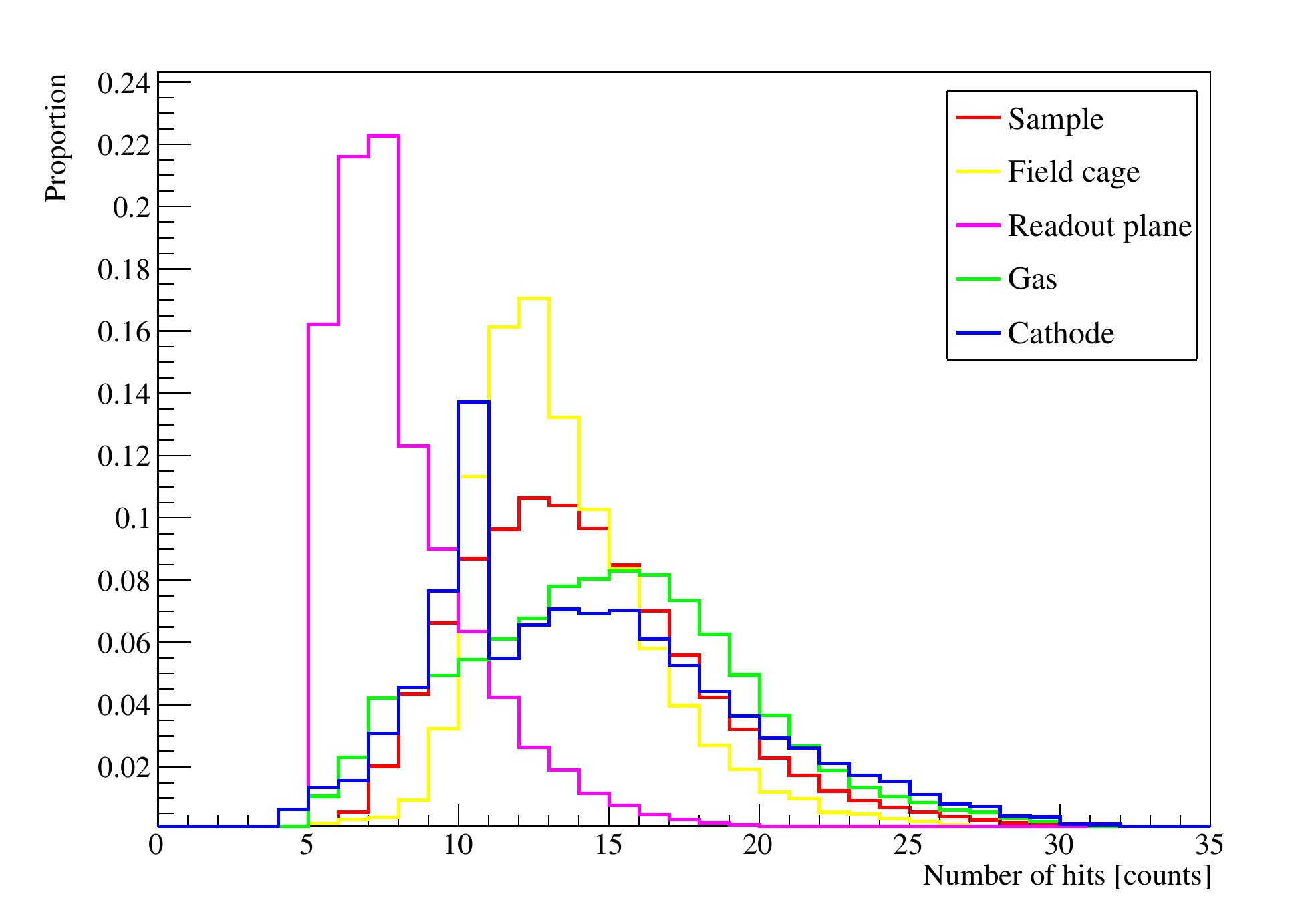}
	\caption{Distribution of number of hits of an event. We created a cut at 10 (9 when $\lambda=1\,\mu$m) to remove the events with short tracks. }
	\label{Nhits}
\end{figure}

We devised two competing algorithms to identify the Bragg peak of a track and thus the track's origin and direction.
A track in the X--Z or Y--Z plane was split into two segments from the mid-point, and the energy in each segment was calculated.
The Bragg peak was more likely to be in the segment with larger energy. 
The energy of each hit point and $dE/dx$ along the vertical direction could also be used to locate Bragg peaks.
When particle energy was higher than 1~MeV and the total number of hits was larger than 10, the identified efficiency of the track's direction was approximately 97\%.

We identified the starting point of each track, and the distribution is shown in Fig.~\ref{Fiducial}. 
The X-axis in the figure is called the fiducial distance, which is defined as the distance in the inward direction from the inner surface of the field cage.
The majority of the events from the field cage and cathode originated from the edge of the sensitive volume.
For those events, the fiducial distance was less than 50~mm.
We performed a fiducial cut at the fiducial distance of 27~mm and discarded any events with a smaller fiducial distance, i.e., closer to the edge of the active volume.
After the fiducial cut, the signal efficiency was at 96.2\%. 
Less than 3\% of the events from field cage and cathode background remained.
The cut was less effective for other background sources, and the efficiencies are listed in Table~\ref{tab:parameter_cut}.

Fig.~\ref{Angle} shows the distribution of $\cos(\theta)$, where $\theta$ is the angle between the particle traveling and upward vertical directions.
The distribution is for events after the fiducial cut.
The majority of $\alpha$ events from the sample had positive $\cos(\theta)$ values while events from the readout plane had negative values, as shown clearly in the figure.
Events from the sample with negative $\cos(\theta)$ values were mostly short events with incorrect track reconstruction.
The same was true for events from the readout plane but with positive $\cos(\theta)$ values.
Events from the cathode and field cage that survived the fiducial cut had incorrectly reconstructed starting points and possibly orientations.
Therefore, the distribution of these events is nearly symmetrical in the figure.
We created a cut at 0 and rejected events with $\cos(\theta)<0$.

We also counted the number of hit strips to further reject background events. 
The distribution of the hits number after the fiducial cut and angle cut is shown in Fig.~\ref{Nhits}.
We maintained only events that triggered more than 10 (9 when $\lambda=1\,\mu$m) strips.
The cut rejected background events from the readout plane most effectively since the remaining events that survived previous cuts were primarily particles with short tracks. 
The peak structure at the number of hits equal to 10 for events from the cathode was an artifact from the angle cut, where we had two algorithms to calculate the angles for tracks longer or shorter than 10 with different efficiencies.

\begin{table}[tb]
 \def\arraystretch{1.2} 
 \tabcolsep 7pt
\caption{The $\alpha$ background rate after each cut ($\lambda=0.1\,\mu$m)}
\label{tab:parameter_cut}
\begin{tabular}{lcccc}
\toprule
Cut parameter &Energy &Fiducial &Angle &Hit number \\
 &cut(\%) &cut(\%) &cut(\%) &cut(\%) \\
\hline 
Field cage & 83.9 &2.9 &0.7 &0.5 \\
Cathode & 79.5 &2.7 &1.5 &1.2 \\
Readout plane & 95.1 &80.3 &7.1 &1.2 \\
Working gas & 92.0 &77.1 &40.5 &28.9 \\
Sample & 98.9 &96.2 &92.2 &74.2 \\ 
\bottomrule
\end{tabular}
\end{table}

\begin{table}[tb]
 \def\arraystretch{1.2} 
 \tabcolsep 7pt 
 \caption{The $\alpha$ background rate after each cut ($\lambda=1\,\mu$m)}
\label{tab:parameter_cut_0.1}
\begin{tabular}{lcccc}
\toprule
Cut parameter &Energy &Fiducial &Angle &Hit number \\
 &cut(\%) &cut(\%) &cut(\%) &cut(\%) \\
\hline 
Field cage & 83.5 &2.8 &0.7 &0.6 \\
Cathode & 78.7 &2.7 &1.3 &0.9 \\
Readout plane & 94.0 &79.3 &10.1 &1.2 \\
Working gas & 92.0 &77.1 &40.5 &31.3 \\
Sample & 97.0 &94.3 &87.5 &67.9 \\ 
\bottomrule
\end{tabular}
\end{table}

\begin{figure}[tb]
	\centering
	\subfigure[Energy spectra after all the cuts assuming $\lambda=0.1\,\mu$m]{
		\includegraphics[width=\columnwidth]{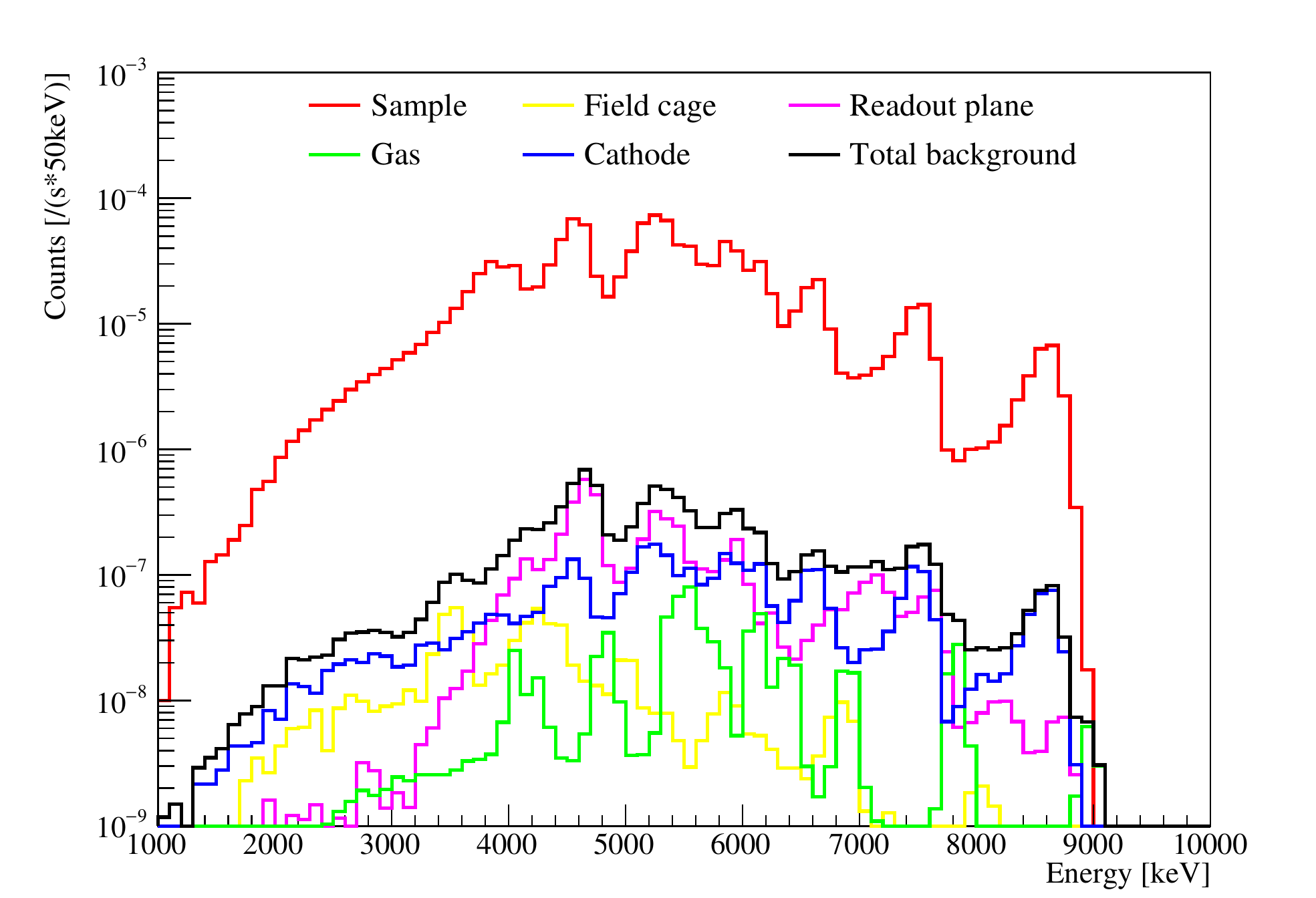}}
	\subfigure[Energy spectra after all the cuts assuming $\lambda=1\,\mu$m]{
		\includegraphics[width=\columnwidth]{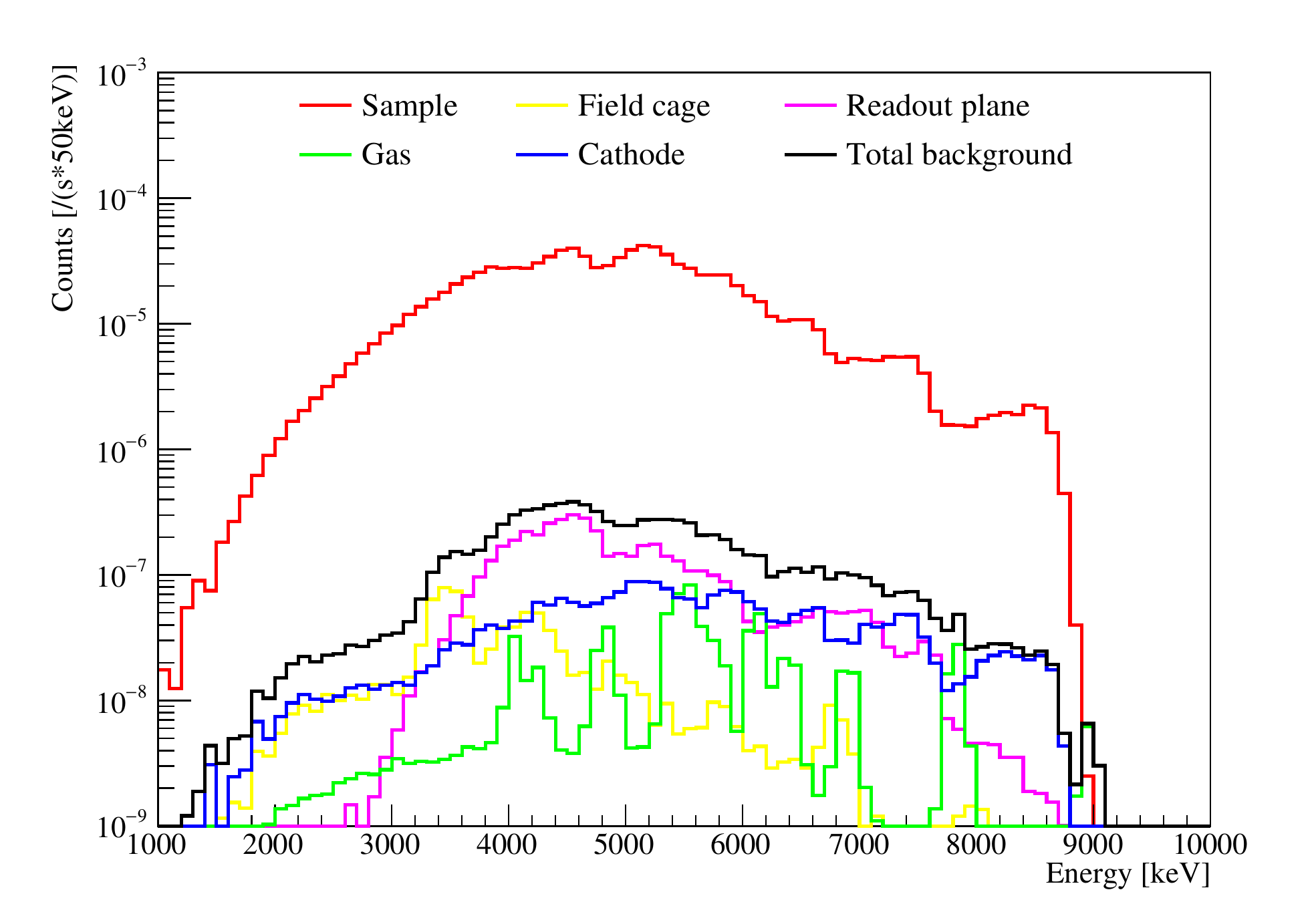}}
	\caption{Energy spectra after all the cuts. The total backgrounds were suppressed by nearly two orders of magnitude while nearly 70\% signals were preserved. }
	\label{reduced_background_spectra}
\end{figure}

Fig.~\ref{reduced_background_spectra} shows the signal and background spectra after all cuts.
The backgrounds were suppressed by nearly two orders of magnitude while maintaining a high signal efficiency.
For $\lambda=0.1\,\mu$m, the combination of all the cuts reduced the $\alpha$ background rate from 82.3 to 0.9 counts per day while maintaining 74.2\% of the signals.
Among the remaining background events, 49.7\% emanated from the readout plane, 36.5\% from the cathode, 7.1\% from the field cage, and 6.7\% from the gas.
Assuming $\lambda=1\,\mu$m, the background was suppressed from 76.1 to 0.8 counts per day while 67.9\% of the signal was preserved.
Approximately 52.5$\%$ of the remaining background was from the readout plane, 28.7$\%$ from the cathode, 10.2\% from the field cage, and 8.6\% from the gas.
Owing to more $\alpha$ events generating below the surface and thus losing partial energy before they were emitted from the surface, the total residual events of both the background and signal were slightly fewer than those of $\lambda=0.1\,\mu$m.

\subsection{Screener3D measurement sensitivity}
\begin{table}[tb]
 \def\arraystretch{1.5} 
 \tabcolsep 7pt 
 
 \caption{Sensitivities of Screener3D for total $\alpha$ measurement (the results in parentheses from $\lambda=1\,\mu$m)}
 \label{tab:Counter_sensitive}
 \begin{tabular}{lcc}
 \toprule
 Measurement time & Background events &Sensitivity \\
 (day) &(counts) &($\mu$Bq$\cdot$m$^{-2}$) \\
 \hline
% 1 &0.93 (0.79) &149 (153) \\
 2 &1.86 (1.59 ) &97 (99) \\
 7 &6.50 (5.56) &46 (47) \\
 \bottomrule
 \end{tabular}
\end{table}

Assuming no events are observed from the samples statistically different from the fluctuation of background, we can evaluate the sensitivity of Screener3D for a particular screening period T in days. 
With a background rate of $R_{bkg}$ counts per day, the total background for a measurement is $B=R_{bkg}T$. 
We use the Poisson fluctuation of small background values to calculate the upper limit on counts and subsequent surface background rate. 
For small $B$ and a confidence level of $\Gamma$, the upper limit for counts of signal $\hat{S}$ can be expressed as 
\begin{equation}
\sum_{k=\hat{S}+B}^{\infty}p_B(k)=1-\Gamma,
\label{Eq:CL}
\end{equation}
in which $\hat{S}$ can be evaluated numerically or with a lower incomplete Gamma function.
Finally, the sensitivity can be calculated using the signal efficiency, $\hat{S}$, and sample surface areas.
Table~\ref{tab:Counter_sensitive} shows the sensitivity of Screener3D with typical measurement periods. 
We can reach a sensitivity of 97~$\mu$Bq$\cdot$m$^{-2}$ at the 90\%confidence level (C.L.) for a two-day measurement, assuming a typical sample such as copper and an exponential surface contamination profile with $\lambda=0.1\,\mu$m.
The sensitivity would be higher than 50~$\mu$Bq$\cdot$m$^{-2}$ when the measurement duration is extended to one week.
For a characteristic depth $\lambda=1\,\mu$m, the results are very similar (Table~\ref{tab:Counter_sensitive}).

\begin{table}[tb] 
 \def\arraystretch{1.5} 
 \tabcolsep 4pt 
 \caption{Sensitivities of Screener3D for $\alpha$ measurement from polonium (the data in parentheses from $\lambda$=1~$\mu$m)}
 \label{tab:sensitive}
 \begin{tabular}{lccc}
 \toprule
 Nuclide &Measurement time &Background events &Sensitivity \\
 &(day) &(counts) &($\mu$Bq$\cdot$m$^{-2}$) \\
 \hline
 $^{212}$Po &2 & 0.06 (0.04) &75 (75) \\
 $^{214}$Po &2 & 0.18 (0.1) &75 (75) \\
 \bottomrule 
 \end{tabular}
\end{table}

Screener3D can also measure $\alpha$ rates of a specific nuclide, such as $^{212}$Po of the thorium decay chain and $^{214}$Po of the uranium decay chain.
The signature $\alpha$ peak of $^{212}$Po is at 8.8~MeV and we define a specific ROI between 8 and 9 MeV.
In the ROI, 7\% of background events and 98\% of the signals are reserved after the energy and topology cuts (fiducial and angle cuts).
The final background rate is 0.03 (0.02 when $\lambda=0.1\,\mu$m) counts per day.
For $^{214}$Po in the ROI of 7 and 8 MeV, 97\% of the background is removed while 98\% of the signals remains.
The final background rate is 0.09 (0.05 when $\lambda=0.1\,\mu$m) counts per day.
For the near-zero background measurements at specific peaks, the Eq.~\ref{Eq:CL} is no longer applicable.
We use the Feldman-Cousins approach~\cite{FC} with zero background as an approximation for this calculation. 
The sensitivity is 75~$\mu$Bq$\cdot$m$^{-2}$ for two-day measurements of both polonium isotopes.

\section{Summary and outlook}
Background events from detector surfaces may have a severe negative impact on experiments searching for rare events.
Screening methods for surface contaminations have not been utilized as widely as those for bulk contaminations (such as gamma and mass spectrometers).
In this paper, we propose a charged-particle detector using the TPC concept for surface contamination screening.
The TPC design concept and detector background control are described in detail.
In addition, Screener3D can significantly reduce $\alpha$ background rate from itself through topological characteristics of particle trajectories, a distinctive feature of the gaseous detector.
Eventually, we can have a detector with fewer than one $\alpha$ background count per day.
With a large active area of approximately 2000~cm$^2$, the TPC we propose may attain a sensitivity higher than 100~$\mu$Bq$\cdot$m$^{-2}$ for surface $\alpha$ contaminations with two days of measurements.
If we only count $\alpha$ particles from specific $\alpha$ peaks, the sensitivity can be further increased.

Screener3D is currently designed primarily for surface $\alpha$ particle screening, and the ROI is typically above 1~MeV.
We can also adopt the detector for surface $\beta$ counting in the energy range below 1~MeV.
However, in this energy range, $\gamma$-rays from the detector and nearby laboratory environment are the dominating background.
Therefore,  a $\gamma$-shielding facility is required and is currently under design.
We are also investigating track reconstruction algorithms to better reconstruct the more meandering $\beta$ tracks, aiming to locate the starting position with high precision. 
Screening capability for surface $\beta$ will be reported in a future manuscript.

\section{Acknowledgements}
This work was supported by the grant from the Ministry of Science and Technology of China (No.~2016YFA0400302) and the grants from the National Natural Sciences Foundation of China (No.~11775142 and No.~U1965201). We appreciate the support from the Chinese Academy of Sciences Center for Excellence in Particle Physics (CCEPP).


\begin{thebibliography}{99} 
\bibitem{DellOro:2016tmg} S. DellOro, S. Marcocci, M. Viel et al, Neutrinoless double beta decay:2015 review. Adv.High Energy Phys {\bf 2016}, 2162659 (2016).
\href{http://dx.doi.org/10.1038/10.1155/2016/2162659}{http://dx.doi.org/10.1038/10.1155/2016/2162659}

\bibitem{Roszkowski:2017nbc} L. Roszkowski, E.M. Sessolo, and S. Trojanowski, WIMP dark matter candidates and searches-current status and future prospects. Rept. Prog. Phys {\bf 81(6)}, 066201 (2018). \href{http://dx.doi.org/10.1088/1361-6633/aab913}{http://dx.doi.org/10.1088/1361-6633/aab913}

\bibitem{Liu:2017drf} J. Liu, X. Chen, and X. Ji, Current status of direct dark matter detection experiments. Nature Phys {\bf 13(3)}, 212–216 (2017). 
\href{http://dx.doi.org/10.1038/nphys4039}{http://dx.doi.org/10.1038/nphys4039}

\bibitem{Cheng:2018lcf} J.-P. Cheng, K.-J. Kang, J.-M. Li et al, The China Jinping Underground Laboratory and its Early Science. Ann. Rev. Nucl. Part. Sci {\bf 67},231–251 (2017). \href{http://dx.doi.org/10.1146/annurev-nucl-102115-044842}{http://dx.doi.org/10.1146/annurev-nucl-102115-044842}

\bibitem{Abgrall:2016cct}N. Abgrall, I.J. Arnquist, F.T. Avignone III, et al, The Majorana Demonstrator radioassay program. Nucl. Instrum. Meth {\bf A, 828}, 22–36 (2016). 
\href{http://dx.doi.org/10.1016/j.nima.2016.04.070}{http://dx.doi.org/10.1016/j.nima.2016.04.070}

\bibitem{Akerib:2014rda} D.S. Akerib, H.M. Araújo, X. Bai, et al, Radiogenic and Muon-Induced Backgrounds in the LUX Dark Matter Detector. Astropart. Phys {\bf 62},33–46 (2015). \href{http://dx.doi.org/10.1016/j.astropartphys.2014.07.009}{http://dx.doi.org/10.1016/j.astropartphys.2014.07.009}

\bibitem{Alessandria:2011vj} F. Alessandria, E.Andreotti, R.Ardito et al, CUORE crystal validation runs: results on radioactive contamination and extrapolation to CUORE background.
Astropart. Phys {\bf 35}, 839–849 (2012). 
\href{http://dx.doi.org/10.1016/j.astropartphys.2012.02.008}{http://dx.doi.org/10.1016/j.astropartphys.2012.02.008}

\bibitem{Jiang:2018pic} H. Jiang, L.P. Jia, Q. Yue, et al, Limits on Light Weakly Interacting Massive Particles from the First 102.8 kg x day Data of the CDEX-10 Experiment. Phys. Rev. Lett {\bf 120(24)},241301 (2018). \href{http://dx.doi.org/10.1103/PhysRevLett.120.241301}{http://dx.doi.org/10.1103/PhysRevLett.120.241301}

\bibitem{Leonard:2007uv} D.S. Leonard, P. Grinberg, P. Weber, et al, Systematic study of trace radioactive impurities in candidate construction materials for EXO-200. Nucl.Instrum. Meth. A {\bf 591}, 490–509 (2008). 
\href{http://dx.doi.org/10.1016/j.nima.2008.03.001}{http://dx.doi.org/10.1016/j.nima.2008.03.001}

\bibitem{Wang:2016eud} X. Wang, X. Chen, C. Fu et al, Material Screening with HPGe Counting Station for PandaX Experiment. JINST {\bf 11(12)},T12002 (2016). 
\href{http://dx.doi.org/10.1088/1748-0221/11/12/T12002}{http://dx.doi.org/10.1088/1748-0221/11/12/T12002}

\bibitem{HPGe_application}X. Guan, L. Ge, G. Zeng, et al. Determination of gross $\alpha$ and $\beta$ activities in Zouma River based on online HPGe gamma measurement system. Nucl. Sci. Tech. {\bf 31(12)},120 (2020). 
\href{http://dx.doi.org/10.1007/s41365-020-00828-0}{http://dx.doi.org/10.1007/s41365-020-00828-0}

\bibitem{ICP-MS_application}L. Yin, Q. Tian, X. Shao, et al. ICP-MS measurement of uranium and thorium contents in minerals in China. Nucl. Sci. Tech. {\bf 27(1)}  10 (2016). 
\href{http://dx.doi.org/10.1007/s41365-016-0018-5}{http://dx.doi.org/10.1007/s41365-016-0018-5}

\bibitem{Alduino:2016vtd} C. Alduino, K. Alfonso, D.R. Artusa, et al, Measurement of the two-neutrino double-beta decay half-life of $^{130}$Te with the CUORE-0 experiment. Eur. Phys. J. C {\bf 77(1)}, 13 (2017). 
\href{http://dx.doi.org/10.1140/epjc/s10052-016-4498-6}{http://dx.doi.org/10.1140/epjc/s10052-016-4498-6}

\bibitem{Zhang:2018xdp} H. Zhang, H.G. Zhang, A. Abdukerim, et al, Dark matter direct search sensitivity of the PandaX-4T experiment. Sci. China Phys. Mech. Astron {\bf 62(3)}, 31011 (2019). \href{http://dx.doi.org/10.1007/s11433-018-9259-0}{http://dx.doi.org/10.1007/s11433-018-9259-0}

\bibitem{OrtecWebsite} Ortec detectors. 
\href{https://www.ortec-online.com/products/radiation-detectors/}{https://www.ortec-online.com/products/radiation-detectors/}

\bibitem{BiPo-3} P.Loaiza, A.S. Barabash, A. Basharina-Freshville, et al, The BiPo-3 detector. Appl. Radiat. Isot{\bf 123}, 28242294(2017). 
\href{https://doi.org/10.1016/j.apradiso.2017.01.021}{https://doi.org/10.1016/j.apradiso.2017.01.021}

\bibitem{BiPo-3_background} A.S. Barabash, A. Basharina-Freshville, E. Birdsall, et al, The BiPo-3 detector for the measurement of ultra low natural radioactivities of thin materials. J. Instrum{\bf12}, P06002(2017).
\href{https://doi.org/10.1088/1748-0221/12/06/p06002}{https://doi.org/10.1088/1748-0221/12/06/p06002}


\bibitem{XIAWebsite} XIA Ultralo 1800 gas counter. 
\href{https://xia.com/ultralo.html}{https://xia.com/ultralo.html}

\bibitem{BetaCage} R. Bunker, et al. The BetaCage, an Ultra-Sensitive Screener for Surface Contamination. AIP Conference Proceedings {\bf 1549}, 132–35 (2013). \href{https://doi.org/10.1063/1.4818093}{https://doi.org/10.1063/1.4818093}

\bibitem{upic} H. Ito, et al, Development of an Alpha-Particle Imaging Detector Based on a Low Radioactive Micro-Time-Projection Chamber. Nucl. Instrum. Meth. A. {\bf 953}, 163050 (2020). \href{https://doi.org/10.1016/j.nima.2019.163050}{https://doi.org/10.1016/j.nima.2019.163050}

\bibitem{MM} S. Andriamonje, D. Attie, E. Berthoumieux at el, Development and performance of Microbulk Micromegas detectors. Journal of Instrumentation {\bf 4}, 02001(2010).
\href{https://doi.org/10.1088/1748-0221/5/02/p02001}{https://doi.org/10.1088/1748-0221/5/02/p02001}

\bibitem{CAT-TPC} L. Yang, J. Xu, Q. Li, et al. Performance of the CAT-TPC based on two-dimensional readout strips. Nucl. Sci. Tech. {\bf 32}, 85 (2021). \href{https://doi.org/10.1007/s41365-021-00919-6}{https://doi.org/10.1007/s41365-021-00919-6}

\bibitem{PandaX-III} X. Chen, F. Fu, J. Galan, et al. PandaX-III: Searching for neutrinoless double beta decay with high pressure 136Xe gas time projection chambers. Sci. China Phys. Mech. Astron. {\bf 60}, 061011 (2017). \href{https://doi.org/10.1007/s11433-017-9028-0}{https://doi.org/10.1007/s11433-017-9028-0}

\bibitem{COMSOLWebsite} COMSOL. 
\href{https://www.comsol.com}{https://www.comsol.com}

\bibitem{slow_control} X. Yan, X. Chen, Y. Chen, et al, Slow Control System for PandaX-III experiment. JINST{\bf 16(05)},T05004 (2021).
\href{https://doi.org/10.1088/1748-0221/16/05/t05004}{https://doi.org/10.1088/1748-0221/16/05/t05004}

\bibitem{MMbackground}J. Castel, I. Coarasa, T. Dafni, et al, Background assessment for the TREX dark matter experiment. Eur. Phys. J. C. {\bf 79}, 782 (2019).
\href{https://doi.org/10.1140/epjc/s10052-019-7282-6}{https://doi.org/10.1140/epjc/s10052-019-7282-6}

\bibitem{Geant4}S. Agostinelli, J. Allison, K. Amako, et al, (GEANT4 Collaboration), GEANT4: A Simulation toolkit. Nucl. Instrum. Meth. A. {\bf 506}, 250 (2003). 
\href{https://doi.org/10.1016/S0168-9002(03)01368-8}{https://doi.org/10.1016/S0168-9002(03)01368-8}

\bibitem{BUDJAS2009755} D. Budjá, A.M. Gangapshev, J. Gasparro, et al, Gamma-ray spectrometry of ultra low levels of radioactivity within the material screening program for the GERDA experiment. Applied Radiation and Isotopes {\bf 67(5)}, 755 – 758(2009). 
\href{http://dx.doi.org/10.1016/j.apradiso.2009.01.019}{http://dx.doi.org/10.1016/j.apradiso.2009.01.019}

\bibitem{Agnes:2015ftt} P. Agnes, L. Agostino, I.F.M. Albuquerque, et al, Results From the First Use of Low Radioactivity Argon in a Dark Matter Search. Phys. Rev. D {\bf 93(8)},081101, (2016). 
\href{http://dx.doi.org/10.1103/PhysRevD.93.081101}{http://dx.doi.org/10.1103/PhysRevD.93.081101}

\bibitem{Simgen} H. Simgen, and G. Zuzel, Analysis of the $^{222}$Rn concentration in argon and a purification technique for gaseous and liquid argon. Appl Radiat Isot. {\bf 67(5)}, 922-5(2009).
\href{http://dx.doi.org/10.1016/j.apradiso.2009.01.058. Epub 2009 Jan 25. PMID: 19251429}{http://dx.doi.org/10.1016/j.apradiso.2009.01.058. Epub 2009 Jan 25. PMID: 19251429}

\bibitem{Juno_acrylic} C.Y. Cao, N. Li, X.Y Yang at el, A practical approach of high precision U and Th concentration measurement in acrylic. Nucl. Instrum. Meth. A {\bf 1004}, 165377(2021).
\href{https://doi.org/10.1016/j.nima.2021.165377}{https://doi.org/10.1016/j.nima.2021.165377}

\bibitem{CUORE_copper} F. Alessandria, R. Ardito, D.R. Artusa at el, Validation of techniques to mitigate copper surface contamination in CUORE. Astroparticle Physics {\bf 45}, 13-22(2013).
\href{https://doi.org/10.1016/j.astropartphys.2013.02.005}{https://doi.org/10.1016/j.astropartphys.2013.02.005}

\bibitem{REST} J. Galan, X. Chen, H. Du, et al, Topological background discrimination in the {PandaX}-{III} neutrinoless double beta decay experiment. J. Phys. G: Nucl. Part. Phys. {\bf 47}, 045108 (2020).
\href{https://doi.org/10.1088/1361-6471/ab4dbe}{https://doi.org/10.1088/1361-6471/ab4dbe}

\href{http://dx.doi.org/10.1007/s11433-017-9028-0}{http://dx.doi.org/10.1007/s11433-017-9028-0}

\bibitem{Garfield} R. Veenhof, Garfield - simulation of gaseous detectors.
\href{http://garfield.web.cern.ch/garfield/}{http://garfield.web.cern.ch/garfield/}

%\bibitem{CUORE_sensitivity} F. Alessandria, R. Ardito, D. R. Artusa, et al, Sensitivity of CUORE to Neutrinoless Double-Beta Decay. arXiv {\bf v3}, (2013). 
%\href{https://arxiv.org/abs/1109.0494v3}{https://arxiv.org/abs/1109.0494v3}
\bibitem{FC}G. J. Feldman and R. D. Cousins, Unified approach to the classical statistical analysis of small signals. Phys. Rev. D {\bf 57}, 3873 (1998).
\href{https://doi.org/10.1103/PhysRevD.57.3873}{https://doi.org/10.1103/PhysRevD.57.3873}
\end{thebibliography}
\end{document}